\documentclass[showpacs,preprintnumbers,amsmath,amssymb,APSl,prd,nofootinbib,superscriptaddress,12pt]{revtex4-2}

\usepackage{hyperref}
\usepackage{amsfonts}
\usepackage{bm}
\usepackage[a4paper, margin=2cm]{geometry}
\usepackage{mathrsfs}
\usepackage{soul}
\usepackage{makecell,multirow}
\usepackage{rotating}
\usepackage[utf8]{inputenc}
\usepackage{amsmath}
\usepackage{makecell}

\usepackage{amssymb}
\usepackage{tensor}
\usepackage{graphicx}
\usepackage{bigints}
\usepackage{bbm}
\usepackage{graphicx}
\usepackage{subfigure}
\usepackage{epsf}
\usepackage{bm}
\usepackage{amsmath}
\usepackage{amsfonts}
\usepackage{amssymb}
\usepackage{graphicx}
\usepackage{tabularx}
\usepackage{multirow}
\usepackage{color}%
\providecommand{\U}[1]{\protect\rule{.1in}{.1in}}

\def\e{\mbox{\rm e}}

\newcommand{\ie}{\begin{equation}}
\newcommand{\fe}{\end{equation}}

\newcommand{\mincir}{\raise
-3.truept\hbox{\rlap{\hbox{$\sim$}}\raise4.truept\hbox{$<$}\ }}
\newcommand{\magcir}{\raise
-3.truept\hbox{\rlap{\hbox{$\sim$}}\raise4.truept\hbox{$>$}\ }}

\providecommand{\U}[1]{\protect\rule{.1in}{.1in}}

\usepackage{tikz,xcolor,hyperref}

\usepackage{hyperref}             
\hypersetup{
    colorlinks=true,              
    breaklinks=true,              
    citecolor=blue,               
    linkcolor=[rgb]{0,0.5,0.9},   
    urlcolor=red,                 
    filecolor=green               
}

\definecolor{lime}{HTML}{A6CE39}
\DeclareRobustCommand{\orcidicon}{%
	\begin{tikzpicture}
	\draw[lime, fill=lime] (0,0) 
	circle [radius=0.16] 
	node[white] {{\fontfamily{qag}\selectfont \tiny ID}};
	\draw[white, fill=white] (-0.0625,0.095) 
	circle [radius=0.007];
	\end{tikzpicture}
	\hspace{-2mm}
}

\foreach \x in {A, ..., Z}{%
	\expandafter\xdef\csname orcid\x\endcsname{\noexpand\href{https://orcid.org/\csname orcidauthor\x\endcsname}{\noexpand\orcidicon}}
}

\begin{document}



\title{The Flight of the Bumblebee in a Non-Commutative Geometry:\\ A New Black Hole Solution}


\author{A. A. Ara\'{u}jo Filho\orcidB{}}
\email{dilto@fisica.ufc.br (The corresponding author)}
\affiliation{Departamento de Física, Universidade Federal da Paraíba, Caixa Postal 5008, 58051-970, João Pessoa, Paraíba, Brazil}
\affiliation{Departamento de Física, Universidade Federal de Campina Grande Caixa Postal 10071, 58429-900 Campina Grande, Paraíba, Brazil.}

\author{N. Heidari\orcidA{}}
\email{heidari.n@gmail.com}
\affiliation{Departamento de Física, Universidade Federal de Campina Grande Caixa Postal 10071, 58429-900 Campina Grande, Paraíba, Brazil.}
\affiliation{Center for Theoretical Physics, Khazar University, 41 Mehseti Street, Baku, AZ-1096, Azerbaijan.}
\affiliation{School of Physics, Damghan University, Damghan, 3671641167, Iran.}


\author{Iarley P. Lobo\orcidD{}}
\email{lobofisica@gmail.com}
\affiliation{Department of Chemistry and Physics, Federal University of Para\'iba, Rodovia BR 079 - km 12, 58397-000 Areia-PB,  Brazil.}
\affiliation{Departamento de Física, Universidade Federal de Campina Grande Caixa Postal 10071, 58429-900 Campina Grande, Paraíba, Brazil.}


\author{Yuxuan Shi\orcidE{}}
\email{shiyx2280771974@gmail.com}
\affiliation{Department of Physics, East China University of Science and Technology, Shanghai 200237, China}


\author{Francisco S. N. Lobo\orcidF{}} \email{fslobo@ciencias.ulisboa.pt}
\affiliation{Instituto de Astrof\'{i}sica e Ci\^{e}ncias do Espa\c{c}o, Faculdade de Ci\^{e}ncias da Universidade de Lisboa, Edifício C8, Campo Grande, P-1749-016 Lisbon, Portugal}
\affiliation{Departamento de F\'{i}sica, Faculdade de Ci\^{e}ncias da Universidade de Lisboa, Edif\'{i}cio C8, Campo Grande, P-1749-016 Lisbon, Portugal}


\begin{abstract}

This paper investigates a new black hole solution within the framework of bumblebee gravity, incorporating non--commutative corrections parameterized by $\Theta$ and implemented through the Moyal twist $\partial_r \wedge \partial_\theta$. Notably, the event horizon remains unaffected by $\Theta$, while the surface gravity becomes ill--defined, in agreement with the behavior previously reported for the non--commutative Schwarzschild black hole \cite{Juric:2025kjl}. The propagation of light is examined by analyzing null geodesics, identifying critical orbits, and determining the resulting black hole shadow. To complement these analyses, we explore gravitational lensing by evaluating the deflection angle in both the weak-- and strong--field regimes. Using these results, constraints are derived for the lensing observables by comparing with the Event Horizon Telescope data for $Sgr A^{*}$ and $M87^{*}$. Finally, we close the analysis by deriving additional constraints from standard Solar System experiments, including Mercury’s orbital precession, gravitational light bending, and time--delay measurements.

\end{abstract}

\maketitle

\tableofcontents


\pagebreak

\section{Introduction}

Within the framework of general relativity, gravitational effects arise from the curvature of spacetime. The mathematical formulation of this curvature is encoded in Einstein’s field equations, whose nonlinear nature renders the search for exact solutions highly restrictive. Only in cases where significant symmetry assumptions are imposed do analytic solutions become available, leaving most realistic configurations inaccessible through exact methods \cite{wald2010general,misner1973gravitation}. To address this limitation, approximation techniques are often employed. A prominent example is the linearization of the field equations in the weak-field regime, which enables a perturbative description of metric fluctuations. Such a framework forms the basis for investigating gravitational--wave phenomena. In black hole physics, it clarifies how small disturbances modify quantum radiation, influence the stability of the spacetime, and regulate the transfer of energy and information between the black hole and its surrounding environment.

Many quantum gravity models argue that spacetime possesses an intrinsic cutoff, preventing resolution beyond a certain limit. This limit, commonly linked to the Planck length, marks the scale where classical geometric descriptions are no longer valid. This conjecture motivates the consideration of spacetime models where the coordinate operators no longer commute. Non--commutative geometry, first introduced in the context of string theory, conduces a systematic framework to realize such modifications \cite{szabo2003quantum,3,szabo2006symmetry}. Its applications extend well beyond gravity, finding a central role in the study of supersymmetric Yang--Mills theories, where it has been employed to address questions of finiteness and consistency \cite{ferrari2003finiteness}. In gravitational theories, non--commutative effects are frequently incorporated through the Seiberg--Witten map, which provides a consistent means of deforming the underlying spacetime symmetries and highlighting the physical consequences of non--commutativity \cite{4}.

Non--commutativity has established itself as a significant theoretical tool for re--examining black hole phenomena, providing perspectives that extend beyond the classical framework of general relativity \cite{karimabadi2020non,heidari2023gravitational,zhao2024quasinormal,modesto2010charged,araujo2024effects,Anacleto:2019tdj,heidari2024exploring,r1,lopez2006towards,mann2011cosmological,Heidari:2025sku,nicolini2009noncommutative,campos2022quasinormal,2,araujo202as5properties}. In this approach, the standard continuum picture of spacetime is replaced by a non--commutative structure at small scales, leading to corrections in the geometry of black hole solutions. A substantial amount of research has investigated the consequences of these deformations for black hole dynamics, with particular attention to evaporation processes and the possible existence of non--classical remnants \cite{myung2007thermodynamics,23araujo2023thermodynamics}.

The thermodynamic description of such systems is also profoundly affected. Non--commutative corrections alter fundamental quantities—including entropy, Hawking temperature, and heat capacity—sometimes introducing qualitatively new behavior that is absent in the commutative case. For instance, several models suggest modified evaporation endpoints, novel stability conditions, and phase structures that depart significantly from those predicted by standard black hole thermodynamics \cite{nozari2006reissner,banerjee2008noncommutative,nozari2007thermodynamics,lopez2006towards,sharif2011thermodynamics}.

An alternative approach to implementing non--commutativity in general relativity was proposed by Nicolini et al. \cite{nicolini2006noncommutative}. Instead of deforming the geometric structure of the Einstein equations directly, their method introduces non--commutative corrections through a modification of the matter sector. In this framework, the singular concept of a point--like mass distribution is replaced by a smooth, smeared profile, thereby embedding non--commutative effects in the stress--energy tensor rather than in the spacetime geometry itself. This reinterpretation has the advantage of regularizing the curvature singularities typically associated with point sources, while still preserving the classical form of Einstein’s equations.  

Two explicit functional forms have been widely employed to model the resulting mass distribution. The first is a Gaussian profile,  
\[
\rho_\Theta(r) = \frac{M}{(4\pi \Theta)^{3/2}} \, e^{-r^2/4\Theta} \, ,
\]  
which ensures exponential suppression at large radii and a smooth core near $r=0$. The second is a Lorentzian distribution,  
\[
\rho_\Theta(r) = \frac{M \sqrt{\Theta}}{\pi^{3/2} \left(r^2 + \pi \Theta\right)^2} \, ,
\]  
whose slower fall--off reflects a different way of smearing the source. Both profiles effectively reflect the idea that spacetime non--commutativity replaces singular point particles with extended matter distributions. 

A distinct line of progress in the study of non--commutative gravity has been achieved by Jurić et al.~\cite{Juric:2025kjl}, who developed refined techniques for generating black hole geometries in gauge--theoretic formulations of the theory. Their work critically re--examined the construction originally put forward by Chaichian et al.~\cite{r1}, demonstrating that the earlier scheme omitted an essential higher--order contribution. The absent term, which involved quadratic combinations of the non--commutative parameters $\Theta^{\mu\nu}$, spin connections, and derivatives of the curvature tensor, was given by 
\ie
-\frac{1}{16}\, \Theta^{\nu \rho}\Theta^{\lambda \mathfrak{t}}
\Big[\tilde{\omega}^{ac}_{\nu}\tilde{\omega}^{cd}_{\lambda}
\big(D_{\mathfrak{t}}R^{d5}_{\rho\mu}+\partial_{\mathfrak{t}}R^{d5}_{\rho\mu}\big)\Big].
\nonumber
\fe

By introducing this additional structure, the framework yields nontrivial modifications to the tetrad fields, which in turn affect the metric sector of the resulting black hole solutions, therefore leading to deviations from the original construction proposed in Ref.~\cite{r1}. Motivated by these modifications, Araújo Filho et al.~\cite{AraujoFilho:2025jcu,araujo2025optical} subsequently formulated a mass--deformed Schwarzschild black hole~\cite{araujo2025geodesics}, along with Kalb--Ramond~\cite{araujo2025optical} and Hayward--type extensions~\cite{heidari2025non}, employing the same technique based on the Moyal twist $\partial_r \wedge \partial_\theta$.

Another fundamental issue in theoretical physics is Lorentz invariance, which stands as one of the most fundamental symmetries in nature. Consequently, it is unsurprising that the majority of gravitational theories incorporate this symmetry, and comparatively little effort has been devoted to exploring the consequences of its potential violation in a gravitational setting.
However, a central challenge in exploring Lorentz symmetry violation in curved spacetime lies in its consistent implementation—a problem that remains largely unresolved in phenomenological studies. Historically, two primary approaches have been adopted. In the explicit scheme, a fixed vector or tensor is directly inserted into the theory from the outset. In contrast, the spontaneous scheme generates the constant vector or tensor dynamically, as the vacuum expectation value of a field. For a comprehensive discussion of these methods for introducing Lorentz symmetry breaking in gravitational contexts, we refer the reader to Ref. \cite{Kostelecky:2003fs}.
In fact, in curved spacetime, the bumblebee model provides a particularly convenient framework for implementing spontaneous Lorentz symmetry breaking. Originally introduced in \cite{Kostelecky:2003fs}, this model realizes the symmetry breaking through a dynamical vector field governed by a nontrivial potential with a continuous set of minima. Initial investigations into the modifications of known gravitational solutions within bumblebee gravity were presented in~\cite{Bertolami:2005bh}. Much work has been invested in exploring solutions in bumblebee gravity, and we refer the reader to following references
\cite{Ovgun:2018xys, Casana:2017jkc,Li:2020dln, Maluf:2020kgf,Ovgun:2018ran,Kanzi:2019gtu,Gullu:2020qzu,Seifert:2009gi,Xu:2022frb,Jha:2020pvk,Filho:2022yrk,Liu:2022dcn,Santos:2014nxm,Kanzi:2021cbg,Ding:2020kfr,Lambiase:2023zeo,Mai:2023ggs,Delhom:2022xfo,Xu:2023xqh,Lessa:2025kln,12araujo2025does,amarilo2024gravitational,heidari2024scattering,AraujoFilho:2024ykw}.

In particular, this study addresses a new proposed black hole configuration in the setting of bumblebee gravity, where non--commutative effects, characterized by the parameter $\Theta$, are incorporated through the Moyal twist $\partial_r \wedge \partial_\theta$. The location of the event horizon is found to be independent of $\Theta$, while the surface gravity becomes undefined, echoing the behavior reported for the non--commutative Schwarzschild case \cite{Juric:2025kjl}. Photon motion is explored by tracing null geodesics, locating the photon sphere, and outlining the corresponding shadow. The analysis is extended to gravitational lensing, where deflection angles are obtained in both weak-- and strong--field approximations and confronted with the Event Horizon Telescope observations of $Sgr A^{*}$ and $M87^{*}$. To complement these results, classical Solar System probes—such as Mercury’s perihelion motion, solar light--bending, and radar echo delays—are employed to set additional limits on the model.



\section{The New Black Hole Solution: via Moyal twist}\label{Sec2}

A recent development in the literature has introduced a breakthrough method for deriving new black hole solutions within the framework of non--commutative 
gauge theory~\cite{Juric:2025kjl}. In this approach, non--commutativity is incorporated through the vierbein formalism. The complete derivation of the 
procedure can be found in Ref.~\cite{Juric:2025kjl}. Here, however, we shall focus on highlighting the distinct aspects encountered therein.

Based on Ref.~\cite{Juric:2025kjl}, in general lines, given a spherically symmetric configuration 
(the ``input metric''), one can construct a new black hole solution (the ``output metric'') within the framework of non--commutative gravity by employing a suitable Moyal twist, as discussed in the introduction. In that work, however, the authors restrict their analysis to the particular case where the temporal and radial components of the input metric are related by $g_{tt} = 1/g_{rr}$. Consequently, in order to address new black hole solutions in the context of bumblebee gravity, certain modifications need to be introduced. In what follows, we shall present only the complementary elements not covered in the aforementioned reference. The adopted coordinate system is $(0,1,2,3) = (t,r,\theta,\phi)$.

We now focus on identifying a suitable gauge-field arrangement that maintains spherical symmetry, constructed in the context of the $\mathrm{SO}(4,1)$ symmetry group~\cite{r6,r1,Juric:2025kjl}:
\ie
e^{0}_{\mu} = \left[A(r),0,0,0,\right], \quad e^{1}_{\mu} =\left[\frac{1}{B(r)}, 0,0,0\right], \quad e^{2}_{\mu} = \left[0, r,0,0\right], \quad e^{3}_{\mu} = \left[0,0,r\, \mathrm{sin}\theta,0\right],
\fe
where $A(r) \equiv \sqrt{g_{tt}}$ and $B(r) \equiv \sqrt{g_{rr}^{-1}}$, with 
$g_{tt}$ and $g_{rr}$ denoting the metric components of the ``input'' metric. 
To proceed, we shall employ the same Moyal twist used in  Refs.~\cite{heidari2025non,araujo2025optical,araujo2025optical}, 
namely $\partial_r \wedge \partial_\theta$, represented by the twist matrix 
for $\Theta^{\mu\nu}$
\begin{equation}
	\Theta^{\mu
		\nu} = \left[\begin{matrix}0 & 0 & 0 & 0\\0 & 0 & \Theta & 0\\0 & - \Theta & 0 & 0\\0 & 0 & 0 & 0\end{matrix}\right].
\end{equation}

Moreover, the nonzero components of non--commutative corrected tetrad field are given by:
\begin{equation}
	\begin{split}
		\hat{e}^0_0 = & \, \,  A{\left(r \right)}+\frac{\Theta^{2} r B^{2}{\left(r \right)} \frac{\mathrm{d}^{3}}{\mathrm{d} r^{3}} A{\left(r \right)}}{4} + \frac{\Theta^{2} r B{\left(r \right)} \frac{\mathrm{d}}{\mathrm{d} r} A{\left(r \right)} \frac{\mathrm{d}^{2}}{\mathrm{d} r^{2}} B{\left(r \right)}}{4} + \Theta^{2} r B{\left(r \right)} \frac{\mathrm{d}^{2}}{\mathrm{d} r^{2}} A{\left(r \right)} \frac{\mathrm{d}}{\mathrm{d} r} B{\left(r \right)} \\
        & + \frac{\Theta^{2} r \frac{\mathrm{d}}{\mathrm{d} r} A{\left(r \right)} \left(\frac{\mathrm{d}}{\mathrm{d} r} B{\left(r \right)}\right)^{2}}{2} + \frac{\Theta^{2} B^{2}{\left(r \right)} \frac{\mathrm{d}^{2}}{\mathrm{d} r^{2}} A{\left(r \right)}}{4} + \frac{\Theta^{2} B{\left(r \right)} \frac{\mathrm{d}}{\mathrm{d} r} A{\left(r \right)} \frac{\mathrm{d}}{\mathrm{d} r} B{\left(r \right)}}{2}, \\
		\hat{e}^1_1 & = \, \, \frac{1}{B{\left(r \right)}} + \frac{\Theta^{2} \frac{\mathrm{d}^{2}}{\mathrm{d} r^{2}} B{\left(r \right)}}{4}, \\
		e^1_2&=- \frac{i \Theta r \frac{\mathrm{d}}{\mathrm{d} r} B{\left(r \right)}}{2} - \frac{i \Theta B{\left(r \right)}}{4},\\
		\hat{e}^2_2&= \, r+ \frac{3 \Theta^{2} r B{\left(r \right)} \frac{\mathrm{d}^{2}}{\mathrm{d} r^{2}} B{\left(r \right)}}{4} + \frac{3 \Theta^{2} r \left(\frac{\mathrm{d}}{\mathrm{d} r} B{\left(r \right)}\right)^{2}}{4} + \frac{\Theta^{2} B{\left(r \right)} \frac{\mathrm{d}}{\mathrm{d} r} B{\left(r \right)}}{2}, \\
		\hat{e}^3_3 = & \, \, r \sin{\left(\theta \right)}+ \frac{\Theta^{2} r B{\left(r \right)} \sin{\left(\theta \right)} \frac{\mathrm{d}^{2}}{\mathrm{d} r^{2}} B{\left(r \right)}}{4} + \frac{\Theta^{2} r \sin{\left(\theta \right)} \left(\frac{\mathrm{d}}{\mathrm{d} r} B{\left(r \right)}\right)^{2}}{2} + \frac{\Theta^{2} B{\left(r \right)} \sin{\left(\theta \right)} \frac{\mathrm{d}}{\mathrm{d} r} B{\left(r \right)}}{2}\\
        & - \frac{\Theta^{2} \sin{\left(\theta \right)} \frac{\mathrm{d}}{\mathrm{d} r} B{\left(r \right)}}{4 B{\left(r \right)}} - \frac{i \Theta \cos{\left(\theta \right)}}{4}. \\
	\end{split}
\end{equation}

At this stage, the final step required to determine the “output” metric must be carried out using the relation presented below
\ie
\label{DefMetTensor}
g_{\mu\nu}\left(x,\Theta\right) = \frac{1}{2} \eta_{{a}{b}}\Bigg[\hat{e}^{{a}}_{\mu}(x,\Theta)\ast\hat{e}^{{b}\star}_{\nu}(x,\Theta)+\hat{e}^{{b}}_{\mu}(x,\Theta)\ast\hat{e}^{{a}\star}_{\nu}(x,\Theta)\Bigg].
\fe
In this context, the symbol $\ast$ represents the standard Moyal star product, widely used within the framework of non--commutative geometry. For the rest of this work, we operate in natural units, choosing $\hbar = c = G = 1$. Under these conventions, the spacetime line element takes the explicit form
\begin{equation}
	\begin{split}
		g_{t t} &= \frac{A(r)}{2} \Bigg\{ - \Theta^{2} r B^{2}{\left(r \right)} \frac{\mathrm{d}^{3}}{ \mathrm{d}^{3}} A{\left(r \right)} - \Theta^{2} r B{\left(r \right)} \frac{\mathrm{d}}{\mathrm{d} r} A{\left(r \right)} \frac{\mathrm{d}^{2}}{\mathrm{d} r^{2}} B{\left(r \right)} - 4 \Theta^{2} r B{\left(r \right)} \frac{\mathrm{d}^{2}}{\mathrm{d} r^{2}} A{\left(r \right)} \frac{\mathrm{d}}{\mathrm{d} r} B{\left(r \right)} \\
        & - 2 \Theta^{2} r \frac{\mathrm{d}}{\mathrm{d} r} A{\left(r \right)} \left(\frac{\mathrm{d}}{\mathrm{d} r} B{\left(r \right)}\right)^{2} - \Theta^{2} B^{2}{\left(r \right)} \frac{\mathrm{d}^{2}}{\mathrm{d} r^{2}} A{\left(r \right)} - 2 \Theta^{2} B{\left(r \right)} \frac{\mathrm{d}}{\mathrm{d} r} A{\left(r \right)} \frac{\mathrm{d}}{\mathrm{d} r} B{\left(r \right)} - 2 A{\left(r \right)} A{\left(r \right)} \Bigg\},
\end{split}
\end{equation}
\begin{equation}
	\begin{split}
		g_{r r} & =  \, \, \frac{1}{B^{2}(r)}+\frac{\Theta^{2}  \frac{\mathrm{d}^{2}}{\mathrm{d} r^{2}} B{\left(r \right)} }{ 2B{\left(r \right)}}  \\
\end{split}
\end{equation}
    \begin{equation}
	\begin{split}
		g_{\theta \theta} &=  r^{2} + \frac{3 \Theta^{2} r^{2} B{\left(r \right)} \frac{\mathrm{d}^{2}}{\mathrm{d} r^{2}} B{\left(r \right)}}{2} + \frac{7 \Theta^{2} r^{2} \left(\frac{\mathrm{d}}{\mathrm{d} r} B{\left(r \right)}\right)^{2}}{4} + \frac{5 \Theta^{2} r B{\left(r \right)} \frac{\mathrm{d}}{\mathrm{d} r} B{\left(r \right)}}{4} + \frac{\Theta^{2} B^{2}{\left(r \right)}}{16}, \\
\end{split}
\end{equation}
    \begin{equation}
	\begin{split}
		g_{\phi \phi} = & \, \,  r^{2} \sin^{2}{\left(\theta \right)}+ \frac{\Theta^{2} r^{2} B{\left(r \right)} \sin^{2}{\left(\theta \right)} \frac{\mathrm{d}^{2}}{\mathrm{d} r^{2}} B{\left(r \right)}}{2} + \Theta^{2} r^{2} \sin^{2}{\left(\theta \right)} \left(\frac{\mathrm{d}}{\mathrm{d} r} B{\left(r \right)}\right)^{2}  \\
        & + \Theta^{2} r B{\left(r \right)} \sin^{2}{\left(\theta \right)} \frac{\mathrm{d}}{\mathrm{d} r} B{\left(r \right)} - \frac{\Theta^{2} r \sin^{2}{\left(\theta \right)} \frac{\mathrm{d}}{\mathrm{d} r} B{\left(r \right)}}{2 B{\left(r \right)}} - \frac{\Theta^{2} \sin^{2}{\left(\theta \right)}}{16} + \frac{5 \Theta^{2}}{16}. \\
	\end{split}
\end{equation}

Therefore, after substituting the metric components of the bumbelbee metric accordingly, we obtain:
\ie
\label{metrictensorss}
\mathrm{d}s^{2} = g_{\mu\nu} \mathrm{d}x^{\mu} \mathrm{d}x^{\nu}   = - \mathrm{A} \left(r,\Theta,\lambda\right) \mathrm{d}t^{2} +  \mathrm{B}\left(r,\Theta,\lambda\right) \mathrm{d}r^{2} + \mathrm{C}\left(r,\Theta,\lambda\right)\mathrm{d}\theta^{2} + \mathrm{D}\left(r,\Theta,\lambda\right) \mathrm{d}\varphi^{2},
\fe
in which
\ie \label{gtt}
\mathrm{A}(r,\Theta,\lambda) = 1 - \frac{2 M}{r} - \frac{ M (11 M-4 r)}{2 (1 + \lambda) r^4}\Theta^2,
\fe
\ie \label{grr}
\mathrm{B}(r,\Theta,\lambda) = \frac{1 + \lambda}{1-\frac{2 M}{r}}+ \frac{ M (3 M-2 r)}{2 r^2 (r-2 M)^2}\Theta^2 ,
\fe
\ie \label{gtheta}
\mathrm{C}(r,\Theta,\lambda) = r^2 + \frac{ \left(64 M^2-32 M r+r^2\right)}{16  r (r-2 M) (1+\lambda)}\Theta^2,
\fe
\ie 
\begin{split}\label{gphi}
 &\mathrm{D}(r,\Theta,\lambda) =   \,\, r^2 \sin ^2 \theta 
  \\
  & +\frac{ \left[\cos (2 \theta ) \left(8 M^2-6 (1 + \lambda ) M r-(1 + \lambda ) r^2\right)-8 M^2+26 (1 + \lambda ) M r-9 (1 + \lambda ) r^2\right]}{32 (1 + \lambda ) r (2 M-r)}\Theta ^2.
\end{split}
\fe

Now, let us examine the main features of the above metric, starting with the event horizon $r_{h}$.
Setting $1/\mathrm{B}(r,\Theta,\lambda) = 0$ yields $r_{h} = 2M$.
In this manner, as in the Schwarzschild case \cite{Juric:2025kjl}, the {event} horizon remains unaffected by either $\lambda$ or $\Theta$ in the bumblebee black hole within the non--commutative gauge theory. {However, for a static spacetime the location of the horizon may also be examined
through the Killing horizon associated with the timelike Killing vector
$\chi^\mu=\partial_t^\mu$. The norm of this vector is
$\chi^\mu\chi_\mu=g_{tt}=-A(r,\Theta,\lambda)$, so the Killing horizon is
determined by the condition
\begin{equation}
A(r,\Theta,\lambda)=0 .
\end{equation}
To verify whether the non–commutative correction shifts the horizon position,
we solve this equation perturbatively. Writing the Killing horizon radius as
\begin{equation}
r_K = 2M + \delta r , \qquad \delta r = \mathcal{O}(\Theta^2),
\end{equation}
and expanding $A(r,\Theta,\lambda)$ to leading nontrivial order in $\Theta$,
one obtains
\begin{equation}
r_K
=
2M + \frac{3\,\Theta^2}{16(1+\lambda)M}
+ \mathcal{O}(\Theta^4).
\end{equation}
Therefore, although the root of $g^{rr}$ remains located at $r=2M$ within the
truncated geometry, the Killing horizon receives a small correction at
quadratic order in the non–commutative parameter. This indicates that the
surface defined by $g^{rr}=0$ does not strictly coincide with the Killing
horizon once the $\Theta^2$ corrections are taken into account.}

Another relevant quantity to analyze is the surface gravity $\kappa(r,\Theta)$, defined as \cite{sedaghatnia2023thermodynamical}
\ie
\begin{split}
\kappa(r,\Theta) & = \frac{1}{{2 \sqrt {{{\mathrm{A}(r,\Theta,\lambda)}}{{\mathrm{B}(r,\Theta,\lambda)}}} }}{\left. {\frac{{\mathrm{d}{{\mathrm{A}(r,\Theta,\lambda)}}}}{{\mathrm{d}r}}} \right|_{r = {r_{h}}}}
\\ & = -\frac{M \left(11 \Theta ^2 M+(\lambda +1) r_{h}^3-3 \Theta ^2 r\right)}{\pi  r_{h}^5 \sqrt{\frac{(\lambda +1) \left(11 \Theta ^2 M^2+4 (\lambda +1) M r_{h}^3-4 \Theta ^2 M r_{h}-2 (\lambda +1) r_{h}^4\right) \left(3 \Theta ^2 M^2-2 M r_{h} \left(\Theta ^2+2 (\lambda +1) r_{h}^2\right)+2 (\lambda +1) r_{h}^4\right)}{r_{h}^6 (r_{h}-2 M)^2}}}.
\end{split}
\fe
Nevertheless, upon substituting $r_{h} = 2M$, the surface gravity becomes 
ambiguous. In other words, for the Moyal twist adopted here, this quantity is 
not well defined. This behavior can already be anticipated from the denominator 
of the preceding expression, namely $r_{h} - 2M$. A similar issue was recently 
reported for the Schwarzschild--like black hole constructed with the same Moyal 
twist~\cite{Juric:2025kjl}. Consequently, the analysis of thermodynamic 
quantities derived from the geometry--such as the Hawking temperature, entropy, 
heat capacity, Gibbs free energy, and particle creation--becomes unfeasible due 
to the ambiguity in $\kappa(r,\Theta)$.

On the other hand, to examine the regularity of the black hole under consideration, let us compute the corresponding Kretschmann scalar $K(r,\Theta)$. The full expression is omitted here due to its considerable length. Evaluating the limit as $r \to 0$, we find
$$
\lim_{r \to 0} K(\Theta) = \frac{1552}{3\,\Theta^4}.
$$
It is worth noting that this result was obtained without expanding the connection or other related quantities, in contrast to the procedure adopted in Ref. \cite{Juric:2025kjl}. Under this complete treatment, the black hole is found to be regular. Moreover, in the limit $r \to 0$, $K(r,\Theta)$ is independent of the Lorentz–violating parameter $\lambda$. A similar behavior was recently reported for a different Lorentz–violating configuration, namely a Kalb–Ramond–like black hole \cite{araujo2025optical}.


\section{The Journey of Light}

This section is devoted to the investigation of light trajectories. We begin by 
considering the null geodesics, deriving the corresponding differential equations 
and solving them numerically to obtain the trajectories. Subsequently, we compute 
the effective potential and, by imposing the appropriate boundary conditions, 
determine the associated photon ring. Finally, based on these results, we analyze 
the resulting black hole shadows.

\subsection{The null geodesics}

This part addresses the study of test particle motion along geodesics within 
the considered spacetime. As the evaluation of the Christoffel symbols constitutes the initial and essential step of the procedure, the analysis begins with their explicit determination:
\ie
\frac{\mathrm{d}^{2}x^{\mu}}{\mathrm{d}\mathfrak{t}^{2}} + \Gamma\indices{^\mu_\alpha_\beta}\frac{\mathrm{d}x^{\alpha}}{\mathrm{d}\mathfrak{t}}\frac{\mathrm{d}x^{\beta}}{\mathrm{d}\mathfrak{t}} = 0. \label{geogeo}
\fe

Within the adopted notation, the symbol $\mathfrak{t}$ represents a generic affine parameter that parametrizes the progression through the particle’s journey. From this framework, a system of four coupled differential equations emerge, each of them taking into account a particular coordinate of spacetime. These equations take the form:
\ie
\frac{\mathrm{d} t^{\prime}}{\mathrm{d} \mathfrak{t}} =  -\frac{4 M t' r' \left(11 \Theta ^2 M+(\lambda +1) r^3-3 \Theta ^2 r\right)}{r \left(-11 \Theta ^2 M^2+4 M r \left(\Theta ^2-(\lambda +1) r^2\right)+2 (\lambda +1) r^4\right)},
\fe
\ie
\begin{split}
\frac{\mathrm{d} r^{\prime}}{\mathrm{d} \mathfrak{t}} = & \frac{1}{8 (\lambda +1) r^3 (2 M-r) \left(3 \Theta ^2 M^2-2 M r \left(\Theta ^2+2 (\lambda +1) r^2\right)+2 (\lambda +1) r^4\right)}   
	\\
& \quad \times \Big\{ 8 (\lambda +1) M r^2 \left(r'\right)^2 \left(6 \Theta ^2 M^2+4 (\lambda +1) M r^3-8 \Theta ^2 M r-2 (\lambda +1) r^4+3 \Theta ^2 r^2\right)   \\
& -16 M (2 M-r)^3 \left(t'\right)^2 \left(11 \Theta ^2 M+(\lambda +1) r^3-3 \Theta ^2 r\right) r^3 \left(\theta '\right)^2
 	 \\
&  +(2 M-r) \left(64 \Theta ^2 M^3+64 M^2 \left((\lambda +1) r^3-\Theta ^2 r\right)+M \left(15 \Theta ^2 r^2-64 (\lambda +1) r^4\right) \right. \\ 
&  \left. +16 (\lambda +1) r^5\right) + 4 r^3 \sin ^2(\theta ) (2 M-r) \left(\varphi '\right)^2 \left(2 \Theta ^2 M^3+2 M^2 \left(8 (\lambda +1) r^3-\Theta ^2 r\right) \right. 
\\ & \left. -(\lambda +1) M r^2 \left(16 r^2-\Theta ^2\right)+4 (\lambda +1) r^5\right)  \Big\},
\end{split}
\fe
\ie
\begin{split}
 \frac{\mathrm{d} \theta^{\prime}}{\mathrm{d} \mathfrak{t}} = & \frac{1}{2 r (r-2 M) \left(64 \Theta ^2 M^2-32 M r \left(\Theta ^2+(\lambda +1) r^2\right)+r^2 \left(\Theta ^2+16 (\lambda +1) r^2\right)\right)} 
 	\\
& \quad \times \Big\{ -4 \theta ' r' \left(64 \Theta ^2 M^3+64 M^2 \left((\lambda +1) r^3-\Theta ^2 r\right)+M \left(15 \Theta ^2 r^2-64 (\lambda +1) r^4\right)+16 (\lambda +1) r^5\right) \\
& +r \sin (2 \theta ) (r-2 M) \left(\varphi '\right)^2 \left(8 \Theta ^2 M^2-2 (\lambda +1) M r \left(3 \Theta ^2+16 r^2\right)+(\lambda +1) r^2 \left(16 r^2-\Theta ^2\right)\right) \Big\},
\end{split}
\fe
and, therefore,
\ie
\begin{split}
\frac{\mathrm{d} \varphi^{\prime}}{\mathrm{d} \mathfrak{t}} = & \frac{1}{\Bar{k}} \Big\{   4 \sin (\theta ) \varphi' \Big[  -4 \sin (\theta ) r' \left(2 \Theta ^2 M^3+2 M^2 \left(8 (\lambda +1) r^3-\Theta ^2 r\right) \right. \\
& \left. -(\lambda +1) M r^2 \left(16 r^2-\Theta ^2\right)+4 (\lambda +1) r^5\right) -r \theta ' \cos (\theta ) (r-2 M) \left(8 \Theta ^2 M^2 \right. \\
& \left. -2 (\lambda +1) M r \left(3 \Theta ^2+16 r^2\right)+(\lambda +1) r^2 \left(16 r^2-\Theta ^2\right)\right)   \Big]  \Big\}  ,
\end{split}
\fe
where
\ie
\begin{split}
 \Bar{k} \equiv & \, r (2 M-r) \left(\Theta ^2 \cos (2 \theta ) \left(8 M^2-6 (\lambda +1) M r-(\lambda +1) r^2\right) \right. \\
& \left. + \left(-8 M^2+26 (\lambda +1) M r-9 (\lambda +1) r^2\right)\Theta ^2 +32 (\lambda +1) r^3 \sin ^2(\theta ) (2 M-r)\right).
\end{split}
\fe

\begin{figure}
    \centering
    \includegraphics[scale=0.643]{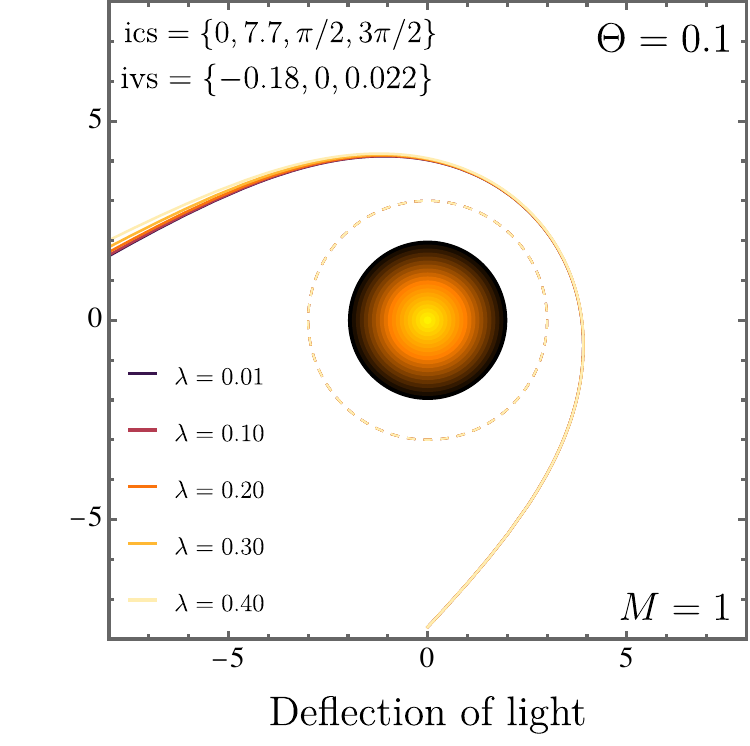}
    \includegraphics[scale=0.643]{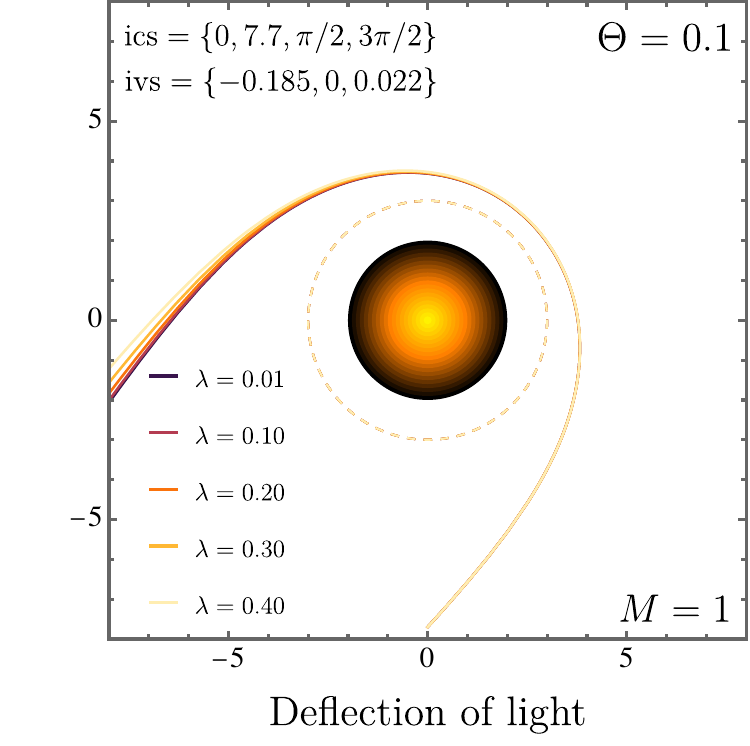}
    \includegraphics[scale=0.643]{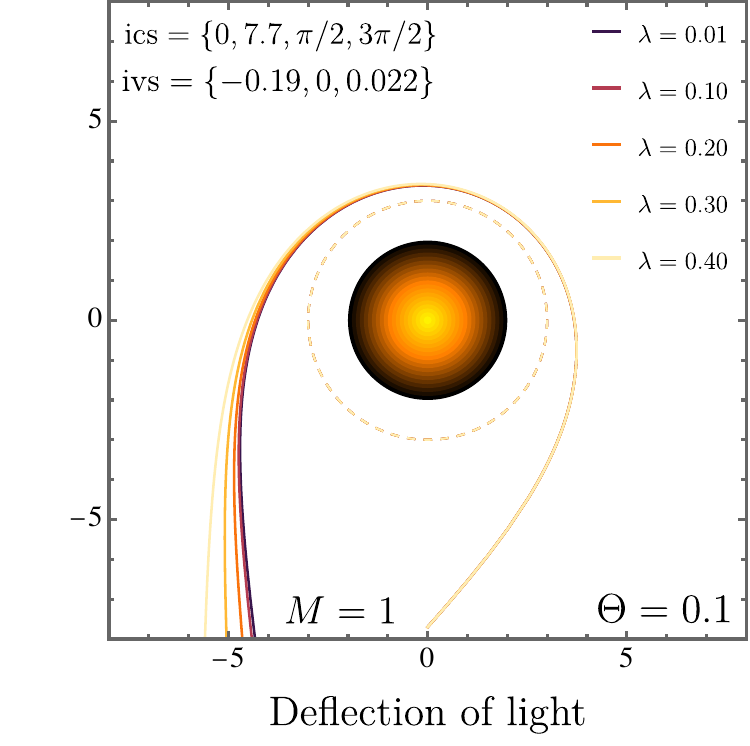}
    \includegraphics[scale=0.643]{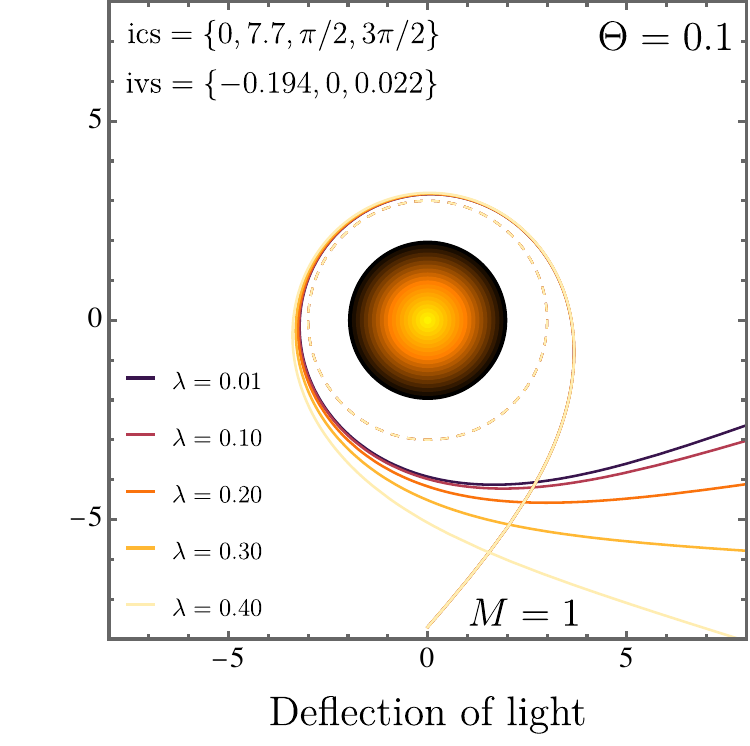}
    \caption{{The null geodesics are displayed for various values of the Lorentz--violating parameter $\lambda$, considering different initial numerical conditions, with $M = 1$ and $\Theta = 0.1$. 
The labels ``ics'' and ``ivs'' shown in the panels denote the initial coordinates and initial velocities adopted in the numerical integration of the geodesic equations, respectively. 
In particular, the initial coordinates correspond to $(t_0,r_0,\theta_0,\phi_0)$, while the initial velocities correspond to $(\dot r_0,\dot\theta_0,\dot\phi_0)$, with the temporal component determined from the null normalization condition. 
The dashed lines represent the associated photon spheres.}}
    \label{trajlight}
\end{figure}

To interpret the lengthy differential equations above, we solve them numerically and present the results in Fig. \ref{trajlight}. The plots show null geodesics for different values of the Lorentz--violating parameter $\lambda$, using various initial numerical conditions, with $M = 1$ and $\Theta = 0.1$. In essence, increasing $\lambda$ causes the photon trajectories to follow more curved paths around the black hole under consideration. Moreover, the corresponding photon spheres are indicated by dashed lines.

\subsection{Critical orbits (photon spheres)}\label{sec:Rc}

The analysis starts from the spacetime geometry given in Eq.~(\ref{metrictensorss}). Employing the Lagrangian approach, we examine the motion of photons, which serves as the basis for describing their trajectories. The corresponding formulation is expressed as
\ie
\mathcal{L} = \frac{1}{2}{g^{(\Theta,\ell)}_{\mu \nu }}{{\dot x}^\mu }{{\dot x}^\nu }
\fe
so that we get
\begin{equation}
\label{lagrangian}
\mathcal{L} = \frac{1}{2}\Big[ - \mathrm{A}(r,\Theta,\lambda){{\dot t}^2} + \mathrm{B}(r,\Theta,\lambda){{\dot r}^2} + \mathrm{C}(r,\Theta,\lambda){{\dot \theta }^2} + \mathrm{D}(r,\Theta,\lambda){{\mathop{\rm \sin}\nolimits} ^2}\, \theta {{\dot \varphi }^2}\Big].
\end{equation}
Let us restrict ourselves to the motion regarding purely the equatorial plane ($\theta = \pi/2$) and employing the Euler–Lagrange framework, the analysis leads to two conserved quantities, namely, energy $E$ and angular momentum $L$. As anticipated, these constants of motion take the form:
\begin{equation}
\label{constant}
E = \mathrm{A}(r,\Theta,\lambda)\,\dot t \quad\mathrm{and}\quad L = \mathrm{D}(r,\Theta,\lambda)\, \dot \varphi.
\end{equation}
For the case of massless modes, we obtain
\begin{equation}\label{light}
- \mathrm{A}(r,\Theta,\lambda){{\dot t}^2} + \mathrm{B}(r,\Theta,\lambda){{\dot r}^2} + \mathrm{D}(r,\Theta,\lambda){{\dot \varphi }^2} = 0.
\end{equation}
Now, let us Substitute Eq.(\ref{constant}) into Eq.(\ref{light}) and carrying out the required algebraic manipulations, we arrive at:
\begin{equation}\label{rdot}
\frac{{{{\dot r}^2}}}{{{{\dot \varphi }^2}}} = {\left(\frac{{\mathrm{d}r}}{{\mathrm{d}\varphi }}\right)^2} = \frac{{\mathrm{D}(r,\Theta,\lambda)}}{{\mathrm{B}(r,\Theta,\lambda)}}\left(\frac{{\mathrm{D}(r,\Theta,\lambda)}}{{\mathrm{A}(r,\Theta,\lambda)}}\frac{{{E^2}}}{{{L^2}}} - 1\right).
\end{equation}
Moreover, it is important to notice that
\ie
\frac{\mathrm{d}r}{\mathrm{d}\Tilde{\lambda}} = \frac{\mathrm{d}r}{\mathrm{d}\varphi} \frac{\mathrm{d}\varphi}{\mathrm{d}\Tilde{\lambda}}  = \frac{\mathrm{d}r}{\mathrm{d}\varphi}\frac{L}{\mathrm{D}(r,\Theta,\lambda)}, 
\fe
with
\ie
\Dot{r}^{2} = \left( \frac{\mathrm{d}r}{\mathrm{d}\Tilde{\lambda}} \right)^{2} =\left( \frac{\mathrm{d}r}{\mathrm{d}\varphi} \right)^{2} \frac{L^{2}}{\mathrm{D}(r,\Theta,\lambda)^{2}}.
\fe
Therefore the effective potential $\mathrm{V}(r,\Theta,\lambda)$ may be expressed as follows:
\ie
\label{effffpotential}
\mathrm{V}(r,\Theta,\lambda) = \frac{{\mathrm{D}(r,\Theta,\lambda)}}{{\mathrm{B}(r,\Theta,\lambda)}}\left(\frac{{\mathrm{D}(r,\Theta,\lambda)}}{{\mathrm{A}(r,\Theta,\lambda)}}\frac{{{E^2}}}{{{L^2}}} - 1\right)\frac{L^{2}}{ \mathrm{D}(r,\Theta,\lambda)^{2}}.
\fe

Having established the necessary preliminaries, the next step in the analysis is to identify the photon spheres, which requires enforcing the condition:
\begin{equation}
\mathrm{V}(r,\Theta,\lambda)=0, \quad\quad \frac{\mathrm{d} \,{\mathrm{V}(r,\Theta,\lambda)}}{\mathrm{d}r} = 0 .
\end{equation}

Here, after taking into account the impact parameter as being $b_c = L/E$, the initial condition $\mathrm{V}(r,\Theta,\lambda) = 0$ yields:
\ie
\label{impactparameter}
b_c=\frac{\mathrm{D}(r,\Theta,\lambda)}{\mathrm{A}(r,\Theta,\lambda)}.
\fe
In this manner, substituting Eq. (\ref{impactparameter}) into Eq. (\ref{effffpotential}) and differentiating with respect to $r$, we get:
\ie
\frac{\mathrm{d} \,{\mathrm{V}(r,\Theta,\lambda)}}{\mathrm{d}r} = \frac{L^2 \left(\mathrm{A}(r,\Theta,\lambda) \mathrm{D}'(r,\Theta,\lambda)-\mathrm{D}(r,\Theta,\lambda) \mathrm{A}'(r,\Theta,\lambda)\right)}{\mathrm{A}(r,\Theta,\lambda) \mathrm{B}(r,\Theta,\lambda) \mathrm{D}(r,\Theta,\lambda)^2}.
\fe
Next, we apply the second condition, $\frac{\mathrm{d} \mathrm{V}(r,\Theta,\lambda)}{\mathrm{d}r} = 0$, to solve for $r$. The resulting solution corresponds to the photon sphere (or critical orbit) and is given by:
\ie \label{eq:rc}
r_{c} = 3 M-\frac{\lambda \Theta^{2}}{9 M}.
\fe

The above expression is obtained up to second order in $\Theta$ and first order in $\lambda$. The first term on the right--hand side coincides precisely with the photon sphere of the bumblebee black hole, which matches the Schwarzschild value. The subsequent term introduces the corrections associated with $\Theta$, meaning that—unlike the pure bumblebee case—non--commutativity allows for modifications of the photon sphere. It is also worth noting that, although the photon spheres of the Kalb--Ramond black and the bumblebee black holes (within a non--commutative gauge theory) are fundamentally different, the second term after the equality is identical to that found in Ref. \cite{araujo2025optical}, which analyzes a non--commutative Kalb--Ramond black hole.

Furthermore, the quantitative values of $r_{c}$ are presented in Tab. \ref{photonspherestab}. In general, keeping $\lambda$ fixed shows that increasing $\Theta$ leads to a decrease in $r_{c}$. A similar feature is seen in the reverse case: fixing $\Theta$ and increasing $\lambda$ also reduces the corresponding photon sphere radius. It is important to note that all these analyses were carried out with $M=1$.

\begin{table}[!ht]
   \centering
    \caption{The critical orbit values are shown by accounting for variations in the non--commutative parameter $\Theta$ and the Lorentz--violating parameter $\lambda$. Here we have considered $M=1$.  } \begin{tabular}{|c||c||c||c||c||c|}
    \hline\hline
         $r_{c}$ & $\lambda = 0.2$ & $\lambda = 0.4$ & $\lambda = 0.6$ & $\lambda = 0.8$& $\lambda = 0.99
        $\\ \hline\hline
        $\Theta= 0.2$ & 2.99911 & 2.99822 & 2.99733 & 2.99644 & 2.99560 \\ \hline
        $\Theta = 0.4$& 2.99644 & 2.99289 & 2.98933 & 2.98578 & 2.98240 \\ \hline
        $\Theta = 0.6$ & 2.99200 & 2.98400 & 2.97600 & 2.96800 & 2.96040 \\ \hline
        $\Theta = 0.8$ & 2.98578 & 2.97156 & 2.95733 & 2.94311 & 2.92960 \\ \hline
         $\Theta = 0.99$ & 2.97822 & 2.95644 & 2.93466 & 2.91288 & 2.89219 \\ \hline
    \end{tabular}
    \label{photonspherestab}
\end{table}


\subsection{Topological photon sphere}

Photon orbits are fundamentally characterized by their stability; they exist in either stable or unstable configurations. The unstable photon orbits are of particular importance, as they govern the formation of the black hole shadow and hence provide an observable signature of the geometry \cite{Vec0-Wei2020,Vec1-Sadeghi2023,Vec5-Cunha2020,Vec4-gashti2025thermodynamic,Vec9-alipour2024weak,araujo2025gravitationalfRT}.

In what follows, we analyze the stability properties of the photon sphere applying the topological method introduced in \cite{Vec0-Wei2020}. This approach relies on defining a regular potential function, given by
\begin{align}\label{eq:Hr}
&H(r, \Theta,\lambda) = \sqrt{\frac{-g_{tt}}{g_{\phi\phi}}}=\sqrt{\frac{\mathrm{A}(r,\Theta,\lambda)}{\mathrm{D}(r,\Theta,\lambda)}}\\ \nonumber
&=\frac{2 \sqrt{2} }{r}\sqrt{1-\frac{2 M}{r}}\Big(11 \Theta ^2 M^2+4 \bar{\lambda} M r^3-4 \Theta ^2 M r-2 \bar{\lambda} r^4\Big)\\ \nonumber
&\Big(4 \sin ^2\theta \left[2 \Theta ^2 M^2-4 \bar{\lambda} M r \left(\Theta ^2+2 r^2\right)+\bar{\lambda} r^2 \left(\Theta ^2+4 r^2\right)\right]-5 \Theta ^2 \bar{\lambda} r \cos ^2\theta (2 M-r)\Big)^{-1},
\end{align}
where $\bar{\lambda}=1+\lambda$. The trigonometric factors $\sin\theta$ and $\cos\theta$ appear naturally and enable a systematic topological characterization of the critical points of the potential.

\begin{figure}[ht!]
\centering
\includegraphics[width=80mm]{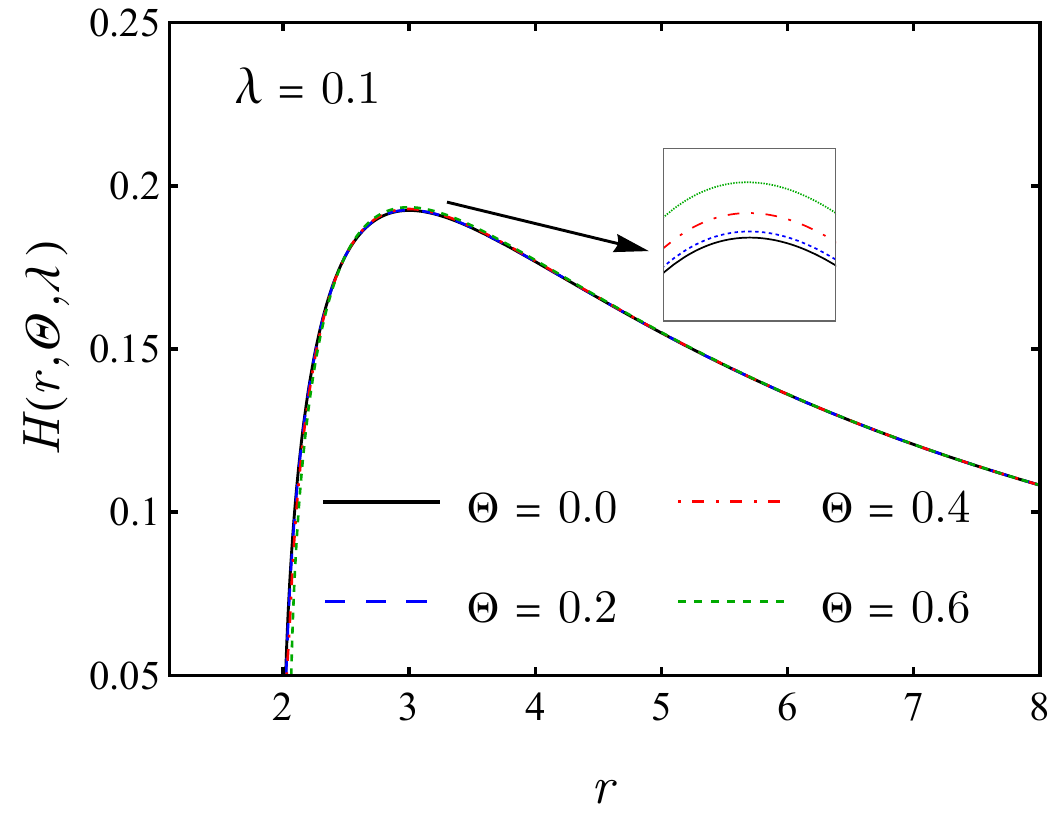}
\caption{Topological potential $H(r,\Theta,\lambda)$ as a function of $r$ for different values of $\Theta$, showing the inward shift of the photon sphere with increasing noncommutativity.}
\label{fig:Hr}
\end{figure}

To investigate the behavior of $H(r,\Theta,\lambda)$, we consider an equatorial plane without loss of generality. If we expand the potential $H(r, \Theta,\lambda)$ up to the second order of $\lambda$ and $\Theta$, the following expression is derived 
\begin{equation}\label{semH}
    H(r,\Theta,\lambda)\simeq \frac{1 }{r}\sqrt{1-\frac{2 M}{r}}\Bigg[1+\left(\frac{\lambda ^2 \left(M r-3 M^2\right)}{r^3 (r-2 M)}+\frac{\lambda  \left(3 M^2-M r\right)}{r^3 (r-2 M)}+\frac{-24 M^2+12 M r-r^2}{8 r^3 (r-2 M)}\right)\Theta ^2 \Bigg].
\end{equation}
According to Eq. \eqref{semH}, the non--commutativity parameter $\Theta$ plays a stronger role than the bumblebee parameter $\lambda$, through coefficient $\Theta^2$. The effect of $\Theta$ on the potential is explored in Fig.~\ref{fig:Hr}. The radius of the photon sphere is derived by finding the extrema of this potential, i.e., by the solutions of $\partial_r H=0$. Holding the mass $M$ and $\lambda$ constant at values of $1$ and $0.5$ respectively, we vary the parameter $\Theta$ through the values $0, 0.2, 0.4$, and $0.6$. The plots indicate that the photon sphere corresponds to a maximum of the potential and is therefore unstable across all values of $\Theta$. This instability implies that even small perturbations of the photon orbit are sufficient to disrupt it, causing the photon to either escape to infinity or be captured by the black hole. 
Furthermore, the location of the maximum shifts slightly toward smaller radii as the non--commutativity parameter $\Theta$ increases. Physically, this means that the non--commutative effects cause the photon sphere to move inward. This result is consistent with the discussion in Sec. \ref{sec:Rc} where the critical orbit expression shows that the non--commutativity decreases the photonic radius (Eq.~\eqref{eq:rc}).

To investigate the photon sphere through a topological framework, we introduce the vector field $\varphi=(\varphi^r,\varphi^\theta)$ with components
\begin{align}
\varphi^r &= \sqrt{-g_{tt}} , \partial_r H(r, \Theta,\lambda) = \sqrt{\mathrm{A}(r, \Theta,\lambda)} , \partial_r H(r, \Theta,\lambda), \label{eq:phi_r} \\
\varphi^\theta &= \frac{1}{\sqrt{g_{\theta \theta}}} , \partial_\theta H(r, \Theta,\lambda) = \frac{1}{\sqrt{\mathrm{C}(r, \Theta,\lambda)}} , \partial_\theta H(r, \Theta,\lambda), \label{eq:phi-theta}
\end{align}
where the vanishing of $\varphi$ signals the location of a photon sphere. For convenience, one may regard this vector field in a complex representation as $\varphi=\varphi^r+i\varphi^\theta$ or
\begin{equation}
\varphi = ||\varphi||\e^{i\theta},\quad\quad\quad\quad||\varphi||=\sqrt{\varphi^i\varphi^i},
\end{equation}
where $i=r~ \text{and }~\theta$. This configuration facilitates the extraction of the topological properties of the vector field. The normalized components of $\varphi$ are defined as
\begin{equation}
n^r = \frac{\varphi^r}{|\varphi|}, \quad n^\theta = \frac{\varphi^\theta}{|\varphi|}.
\label{eq:normalized}
\end{equation}

The normalized components $(n^r, n^\theta)$ serve as a useful tool for tracking how vector directions evolve, making it possible to pinpoint topological phase transitions in the underlying spacetime geometry. In this description, the photon sphere appears as a topological defect located at points where $\varphi$ vanishes. Following the classification proposed in \cite{Vec5-Cunha2020,Vec3-Duane1984}, the winding number of the normalized field along a closed contour $C$ encircling such a zero determines the associated topological charge. Each photon sphere is therefore characterized by a discrete charge $Q=\pm 1$. The $Q=-1$ corresponds to an unstable photon orbit, while $Q=+1$ signals stability.

\begin{figure}[ht!]
\centering
\includegraphics[width=85mm]{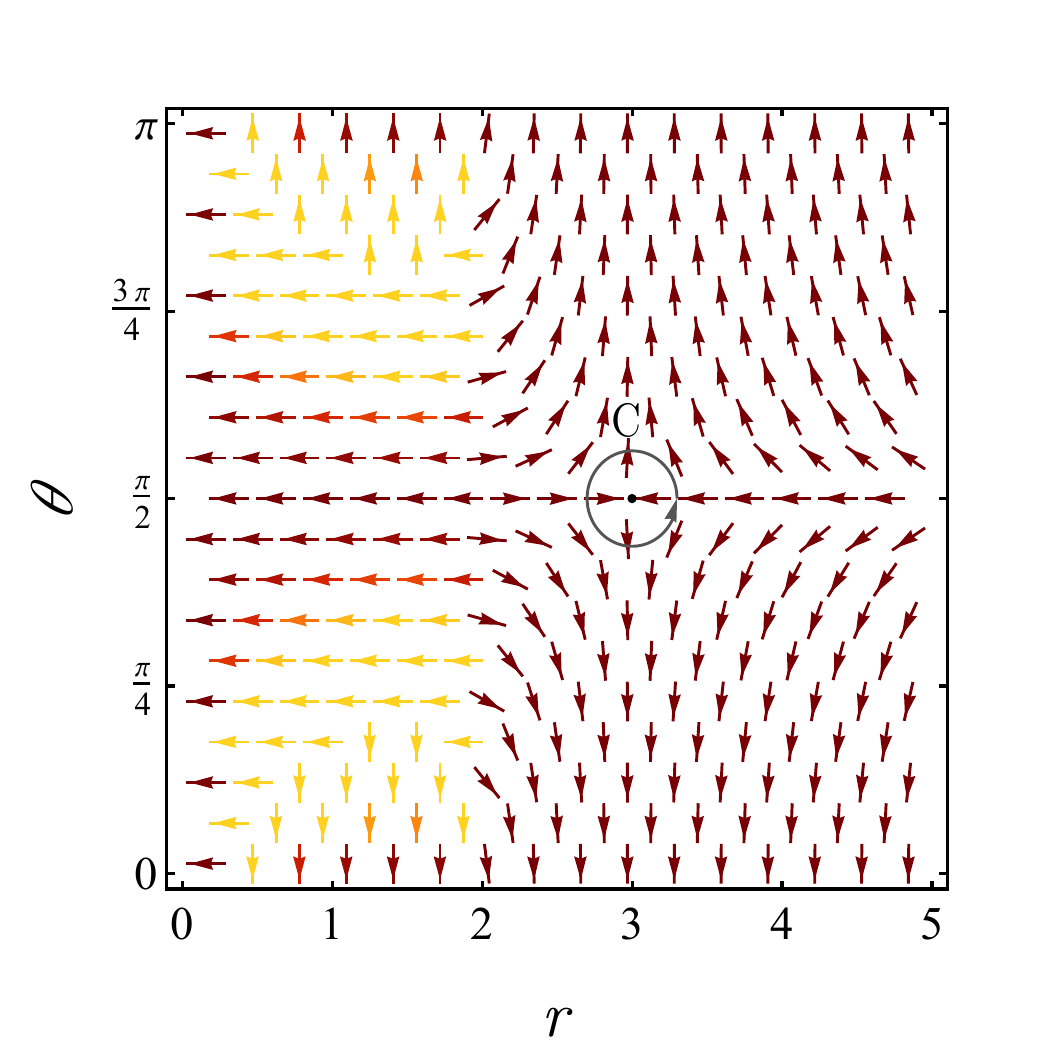}
\caption{Normalized vector field $(n^r, n^\theta)$ in the $(r,\theta)$ plane for $M=1$, $\lambda=0.5$, and $\Theta=0.1$. The closed contour $C$ surrounds the photon sphere at $(r_{\text{ph}},\theta)=(2.99963,\pi/2)$, highlighting the critical point where the topological charge is determined.}
\label{fig:vector}
\end{figure}

The normalized vector field is shown in Fig.~\ref{fig:vector}. In this case, we identify a single photon sphere outside the event horizon, situated at $r_{\text{ph}}=2.99963$. The contour $C$ encloses this critical point with a topological charge of $Q=-1$, thereby confirming that the orbit is unstable. 
In other words, this analysis indicates that no stable circular photon orbits exist in the present background.


\subsection{Shadows}

Now, in order to corroborate our previous findings concerning the critical orbits, let us apply them to the shadows. As it is well known in the literature, the shadow radius can be written as \cite{jusufi2024charged,araujo2024effects,heidari2025absorption,heidari2024impact}: 
\ie
\begin{split}
\label{RRRRR}
\mathcal{R} = & \sqrt {\frac{{{\mathrm{D}(r,\Theta,\lambda)}}}{{{\mathrm{A}(r,\Theta,\lambda)}}}}\Bigg|_{r = r_{c}} = \sqrt{\frac{\frac{\Theta ^2 \left(-16 M^2+32 (\lambda +1) M r_{c}-8 (\lambda +1) r_{c}^2\right)}{32 (\lambda +1) r_{c} (2 M-r_{c})}+r_{c}^2}{-\frac{\Theta ^2 M (11 M-4 r_{c})}{2 (\lambda +1) r_{c}^4}-\frac{2 M}{r_{c}}+1}} \\
& \approx \, \frac{\Theta ^2 \left(24 M^2-12 M r_{c}+r_{c}^2\right)}{8 \sqrt{-\frac{r_{c}^3}{2 M-r_{c}}} (r_{c}-2 M)^2}+\sqrt{-\frac{r_{c}^3}{2 M-r_{c}}} + \left( \frac{\Theta ^2 M (r_{c}-3 M)}{\sqrt{\frac{r_{c}^3}{r_{c}-2 M}} (r_{c}-2 M)^2}  \right)\lambda \\
& \approx \, \, 3 \sqrt{3} M - \frac{\Theta ^2}{8 \left(\sqrt{3} M\right)} + \mathcal{O}(\Theta^{4})\lambda + ...\, .
\end{split}
\fe
Therefore, up to second order in $\Theta$ and first order in $\lambda$, although the Lorentz--violating parameter affects the photon sphere, it does not contribute to the shadow radius. To facilitate the interpretation of our results, we present Tab. \ref{shadoowsstab}. Overall, as $\Theta$ increases, $\mathcal{R}$ decreases. Such a behavior is expected due to the second term of the bottom line in Eq. (\ref{RRRRR}), where $\Theta$ appears with a minus sign, thereby naturally reducing $\mathcal{R}$.

\begin{table}[!ht]
   \centering
    \caption{The shadow radii values are exhibited by taking into account the variations in the non--commutative parameter $\Theta$. Here, it is considered $M=1$.  } \begin{tabular}{|c||c|}
    \hline\hline
         $\mathcal{R}$ & $\text{------}$ 
        \\ \hline\hline
        $\Theta= 0.0$ & 5.19615  \\ \hline
        $\Theta= 0.1$ & 5.19543  \\ \hline
        $\Theta= 0.2$ & 5.19327  \\ \hline
        $\Theta = 0.3$& 5.18966  \\ \hline
        $\Theta = 0.4$ & 5.18461  \\ \hline
        $\Theta = 0.5$ & 5.17811 \\ \hline
         $\Theta = 0.99$ & 5.12542 \\ \hline
    \end{tabular}
    \label{shadoowsstab}
\end{table}


\section{Lensing Phenomena: Weak Field Regime}

In this section, we examine the stability of the critical orbits $r_{c}$ obtained previously. The central question is whether $r_{c}$ is stable or unstable. To address this, we compute the Gaussian curvature and analyze its sign. According to the Cartan--Hadamard theorem, the sign of the curvature can be linked to the presence or absence of conjugate points. In other words, as we shall see in what follows, such a theorem states that on a complete, simply connected Riemannian manifold, a negative Gaussian curvature ($\mathcal{K}<0$) everywhere implies the absence of conjugate points. In contrast, a positive curvature ($\mathcal{K}>0$) can permit their existence. Moreover, in the context of the photon spheres, a negative Gaussian curvature essentially indicates that the orbits are unstable.


\subsection{Stability of $r_{c}$}

Light propagation in the vicinity of black holes is dictated by the structure of the optical manifold, whose properties determine how photons move in curved spacetime. A fundamental aspect of this geometry lies in the occurrence of conjugate points, which provide information about the possible configurations of null geodesics. Whether a photon sphere admits stable or unstable trajectories becomes evident when small deviations from the geodesic are considered. If the configuration is unstable, even a minimal perturbation, the photon either to spiral into the horizon or to escape to infinity. On the other hand, when stability holds, these perturbations do not destroy the orbit but instead allow the photon to remain trapped in bounded paths encircling the black hole \cite{qiao2022curvatures,qiao2022geometric}.

The characterization of photon orbits in a curved spacetime depends strongly on whether conjugate points appear in the manifold. When such points are present, the photon sphere admits stable configurations; their absence, however, marks an unstable regime. This classification can be established through the Cartan–Hadamard theorem, which links the sign of the Gaussian curvature $\mathcal{K}(r)$ to the possibility of conjugate points. In this manner, the curvature itself becomes a purely geometric tool for testing the stability of null critical paths \cite{qiao2024existence}. Following this reasoning, one may represent the null geodesics—trajectories satisfying $\mathrm{d}s^{2}=0$—in the form \cite{araujo2024effects,heidari2025absorption,araujo2025optical}:
\ie
\mathrm{d}t^2=\gamma_{ij}\mathrm{d}x^i \mathrm{d}x^j = \frac{\mathrm{B}(r,\Theta,\lambda)}{\mathrm{A}(r,\Theta,\lambda)}\mathrm{d}r^2  +\frac{\mathrm{\Bar{D}}(r,\Theta,\lambda)}{\mathrm{A}(r,\Theta,\lambda)}\mathrm{d}\varphi^2 .
\fe
In this setting, the indices $i$ and $j$ take values from $1$ to $3$, while $\gamma_{ij}$ designates the corresponding elements of the optical metric. The quantity $\mathrm{\Bar{D}}(r,\Theta,\lambda)$ represents the function $\mathrm{D}(r,\Theta,\lambda)$ restricted to the equatorial section of the geometry, namely at $\theta=\pi/2$. The formulation of the Gaussian curvature associated with this optical manifold, which is essential for the stability analysis, can be found in Ref. \cite{qiao2024existence}:
\ie
\mathcal{K}(r,\Theta,\lambda) = \frac{R}{2} =  - \frac{\mathrm{A}(r,\Theta,\lambda)}{\sqrt{\mathrm{B}(r,\Theta,\lambda) \,  \mathrm{\Bar{D}}(r,\Theta,\lambda)}}  \frac{\partial}{\partial r} \left[  \frac{\mathrm{A}(r,\Theta,\lambda)}{2 \sqrt{\mathrm{B}(r,\Theta,\lambda) \, \mathrm{\Bar{D}}(r,\Theta,\lambda) }}   \frac{\partial}{\partial r} \left(   \frac{\mathrm{\Bar{D}}(r,\Theta,\lambda)}{\mathrm{A}(r,\Theta,\lambda)}    \right)    \right].
\fe

When the parameters $\lambda$ and $\Theta$ are assumed to be small, the two--dimensional Ricci scalar $R$ admits a tractable expansion. Restricting the analysis to terms linear in $\lambda$ and quadratic in $\Theta$, the curvature takes a simplified analytic form. In this approximation, one arrives at:
\ie
\begin{split}
\label{gaussiancurvature}
\mathcal{K}(r,\Theta,\lambda)    \, \,  \approx & \, \frac{3 M^2}{r^4}-\frac{2 M}{r^3} + \frac{2 \lambda  M}{r^3}-\frac{3 \lambda  M^2}{r^4} +\frac{624 \Theta ^2 M^4}{8 M r^7-4 r^8}-\frac{848 \Theta ^2 M^3 r}{8 M r^7-4 r^8}\\
& + \frac{374 \Theta ^2 M^2 r^2}{8 M r^7-4 r^8}+\frac{\Theta ^2 r^4}{8 M r^7-4 r^8}-\frac{54 \Theta ^2 M r^3}{8 M r^7-4 r^8}  +\frac{312 \Theta ^2 \lambda  M^4}{r^7 (r-2 M)}\\
& -\frac{420 \Theta ^2 \lambda  M^3}{r^6 (r-2 M)}+\frac{181 \Theta ^2 \lambda  M^2}{r^5 (r-2 M)} + \frac{\Theta ^2 \lambda }{4 r^3 (r-2 M)}-\frac{24 \Theta ^2 \lambda  M}{r^4 (r-2 M)}.
\end{split}
\fe
The decomposition of the Gaussian curvature can be interpreted by tracing each contribution back to the underlying geometry. The first two terms reproduce the well–known Schwarzschild result, reflecting the standard spherically symmetric structure without any additional fields. The next pair of contributions emerges from the Lorentz–violating modifications introduced by bumblebee gravity. Beyond these, the remaining parts of the expression originate from non–commutative effects as well as from cross–interactions between the bumblebee sector and the non–commutative corrections.

\begin{figure}
    \centering
    \includegraphics[scale=0.56]{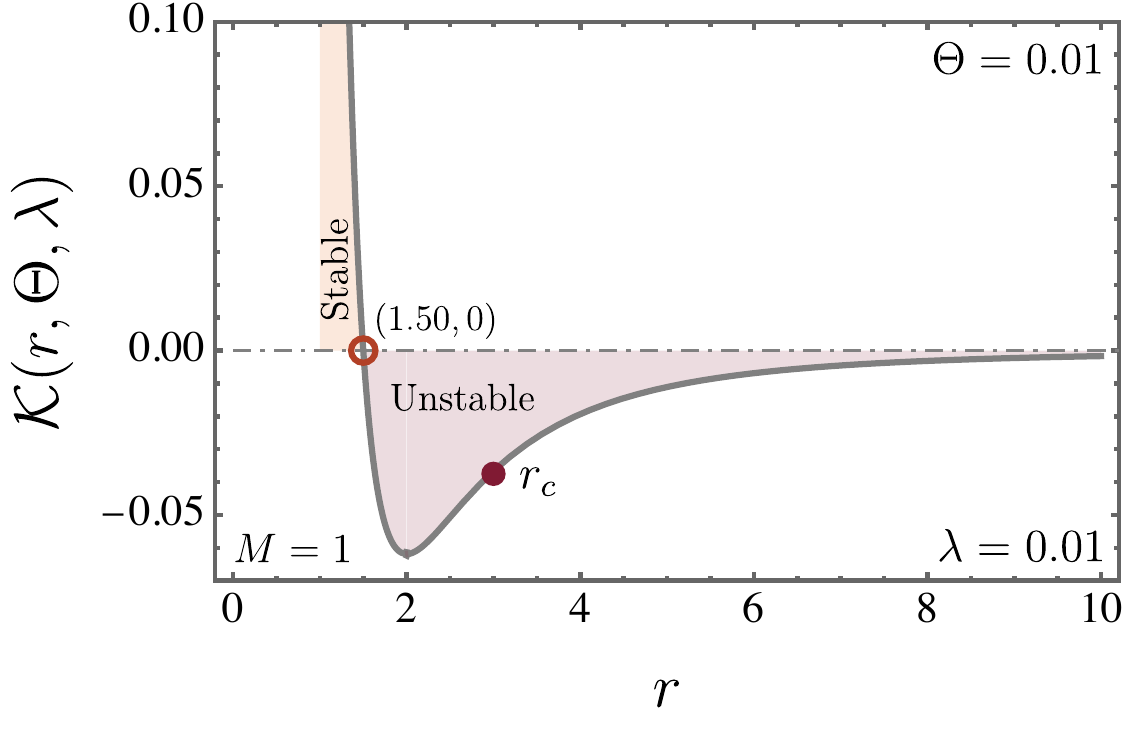}
    \caption{The Gaussian curvature is displayed for the parameter set $M=1$, $\Theta=0.01$, and $\lambda=0.01$. The plot highlights the regions corresponding to stable and unstable photon spheres (or critical orbits). The transition point, where stability changes to instability, is marked with a red circle (at $(1.50,0)$). In addition, the photon sphere radius is represented by a wine--colored dot.}
    \label{krtl}
\end{figure}

The link between photon orbit stability and the curvature of the optical manifold can be formulated through the Gaussian curvature $\mathcal{K}(r,\Theta,\lambda)$. As discussed in Refs. \cite{qiao2022curvatures,qiao2022geometric,qiao2024existence}, the criterion is straightforward: negative values of $\mathcal{K}$ correspond to unstable photon trajectories, while positive values characterize stable photon spheres. This criterion is illustrated in Fig. \ref{krtl}, where $\mathcal{K}(r,\Theta,\lambda)$ is plotted against the radial coordinate $r$ for the parameters $M=1.0$, $\lambda=0.01$, and $\Theta=0.01$. The plot separates regions of stability from instability, with the transition occurring precisely at the point where the curvature flips its sign. In this case, the turning point lies at $(1.50,0)$, highlighted with a wine--colored circle. For the chosen parameters, the photon sphere radius is approximately $r_{ph}\approx2.99999$ (wine dot). Since this radius falls within the region where $\mathcal{K}(r,\Theta,\lambda)<0$, the corresponding null orbits are identified as unstable.

\subsection{The deflection angle}

Beginning with the expression of the Gaussian curvature in Eq. (\ref{gaussiancurvature}), the next step is to employ the Gauss--Bonnet theorem in order to evaluate the deflection angle within the weak–field regime \cite{Gibbons:2008rj}. This procedure demands the computation of the optical surface integral, restricted to the equatorial plane. The corresponding formulation is given by:
\ie
\mathrm{d}S = \sqrt{\gamma} \, \mathrm{d} r \mathrm{d}\varphi = \sqrt{\frac{\mathrm{B}(r,\Theta,\lambda)}{\mathrm{A}(r,\Theta,\lambda)}  \frac{\mathrm{D}(r,\Theta,\lambda)}{\mathrm{A}(r,\Theta,\lambda)} } \, \mathrm{d} r \mathrm{d}\varphi,
\fe
which ultimately yields the expression for the deflection angle, written as
\ie
\begin{split}
\label{deflectionangleweak}
& \Tilde{\alpha} (b,\Theta,\lambda) =  - \int \int_{D} \mathcal{K}(r,\Theta,\lambda)\,  \mathrm{d}S = - \int^{\pi}_{0} \int^{\infty}_{{\big(\frac{\sin (\varphi )}{b}+\frac{M (1-\cos (\varphi ))^2}{ b^2}\big)^{-1}}} \mathcal{K}(r,\Theta,\lambda) \, \mathrm{d}S \\
 \simeq  & \, \, \frac{4 M}{b}+ \frac{ {15} \pi  M^2}{4 b^2} -\frac{ {15} \pi  \lambda  M^2}{8 b^2}  -\frac{2 \lambda  M}{b} {-} \frac{ {587} \pi  \Theta ^2 M^2}{128 b^4}-\frac{ {14} \Theta ^2 M}{3 b^3}+\frac{\pi  \Theta ^2}{16 b^2} \\
 & {+}\frac{ {1547} \pi  \Theta ^2 \lambda  M^2}{256 b^4}+\frac{{7} \Theta ^2 \lambda  M}{3 b^3}-\frac{\pi  \Theta ^2 \lambda }{32 b^2}.
\end{split}
\fe

\begin{figure}
    \centering
    \includegraphics[scale=0.5]{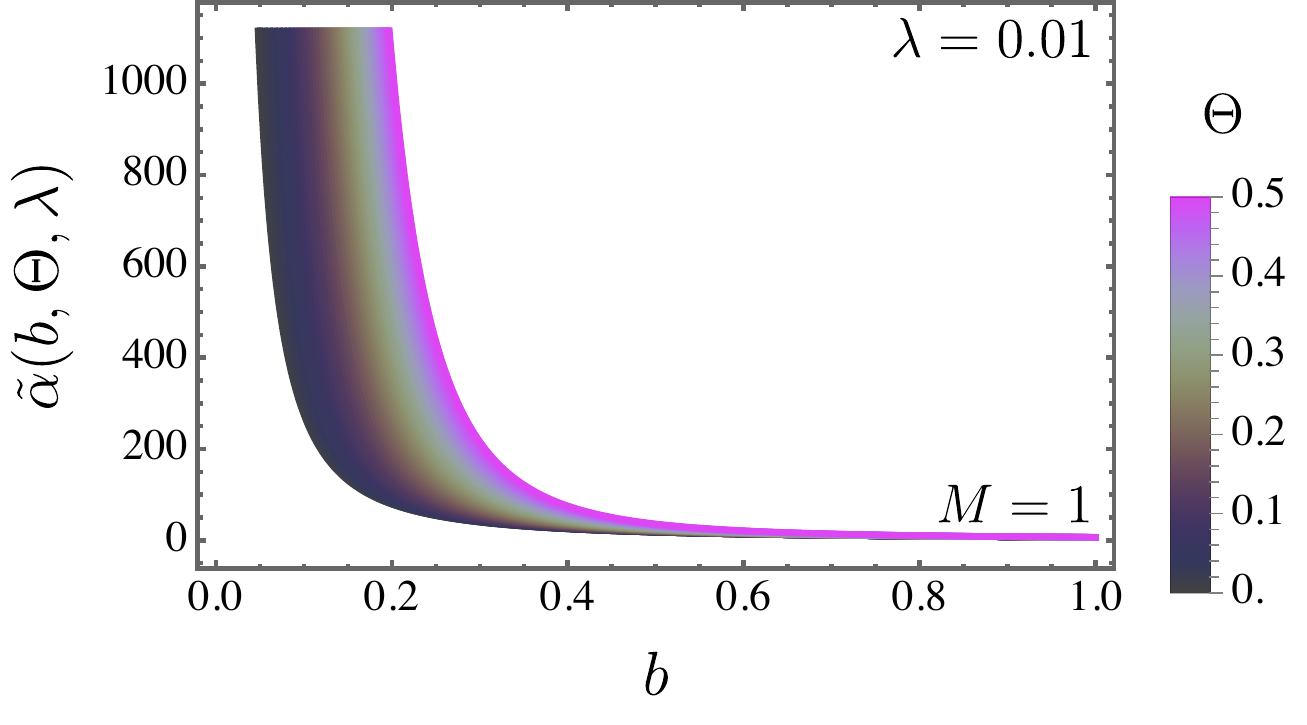}
    \includegraphics[scale=0.5]{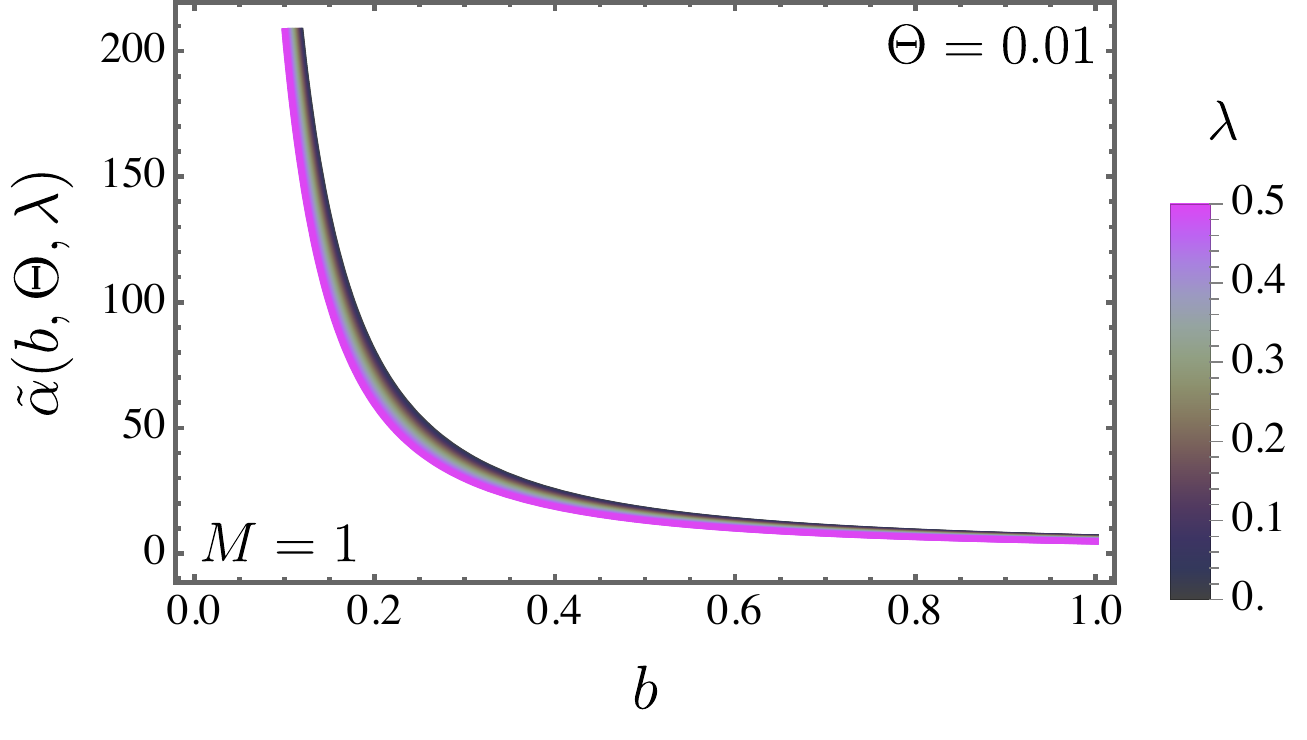}
    \caption{The deflection angle in the weak–field regime, $\Tilde{\alpha}(b,\Theta,\lambda)$, is displayed for two distinct parameter configurations. In the upper panel, the case with $M=1$ and $\lambda=0.01$ is considered while $\Theta$ is varied. The lower panel instead fixes $M=1$ and $\Theta=0.01$, showing how the deflection responds to changes in $\lambda$. }
    \label{deflec}
\end{figure}

The expansion of the weak--field deflection angle can be separated into distinct contributions, each associated with a different physical origin. The initial pair of terms reproduces the standard Schwarzschild result, while the following two reflect corrections induced by the bumblebee background. The next three terms correspond to modifications arising from the non--commutative extension of the Schwarzschild solution. Finally, the last three contributions (shown on the bottom line of Eq. (\ref{deflectionangleweak})) emerge from the interplay between the bumblebee sector and the non--commutative effects.
Fig. \ref{deflec} illustrates how the deflection angle $\Tilde{\alpha}(b,\Theta,\lambda)$ responds to variations in the deformation parameters. For a chosen impact parameter—for instance, $b=0.3$—an increase in $\Theta$ produces a stronger deflection. On the other hand, raising the value of $\lambda$ reduces the bending, thereby lowering the overall magnitude of $\Tilde{\alpha}(b,\Theta,\lambda)$.

\section{Lensing Phenomena: Strong Field Regime}

The following equations of motion are obtained by using Eq. \eqref{lagrangian} and taking into account the null geodesic condition $g_{\mu\nu}\dot{x}^{\mu}\dot{x}^{\nu} = 0$:
\begin{align}
\label{dot_t}
\dot{t} &= -\frac{1}{b \mathrm{A}(r,\Theta,\lambda)}, \\
\label{dot_p}
\dot{\varphi} &= \pm\frac{1}{\mathrm{D}(r,\Theta,\lambda)}, \\
\label{dot_r}
\dot{r}^2 &= -\frac{1}{b^2 \mathrm{A}(r,\Theta,\lambda)\mathrm{B}(r,\Theta,\lambda)} - \frac{1}{\mathrm{B}(r,\Theta,\lambda)\mathrm{D}(r,\Theta,\lambda)}.
\end{align}
The affine parameter $\chi$ has been rescaled to $\chi/|L|$, the impact parameter is specified as $b = |L|/E$, and the $\pm$ sign denotes whether the light beam is moving clockwise or anticlockwise. The orbital equation is obtained by rearranging Eq. \eqref{dot_r}:
\begin{align}
\label{geo1}
\dot{r}^2 + \frac{1}{\mathrm{B}(r,\Theta,\lambda)\mathrm{D}(r,\Theta,\lambda)} = \frac{1}{b^2 \mathrm{A}(r,\Theta,\lambda)\mathrm{B}(r,\Theta,\lambda)}.
\end{align}

In the equatorial plane, for circular photon orbits where $\dot{r} = 0$ and $\ddot{r} = 0$, Eq. \eqref{geo1} simplifies to:
\begin{align}
\label{geo1_re1}
\frac{1}{b^2} = \frac{\mathrm{A}(r,\Theta,\lambda)}{\mathrm{D}(r,\Theta,\lambda)},
\end{align}
and the stability condition requires:
\begin{align}
\label{geo1_re2}
\frac{\mathrm{d}}{\mathrm{d}r}\left(\frac{1}{\mathrm{B}(r,\Theta,\lambda)\mathrm{D}(r,\Theta,\lambda)}\right) = \frac{1}{b^2} \frac{\mathrm{d}}{\mathrm{d}r}\left(\frac{1}{\mathrm{A}(r,\Theta,\lambda)\mathrm{B}(r,\Theta,\lambda)}\right).
\end{align}
The largest real solution to Eqs. \eqref{geo1_re1} and \eqref{geo1_re2} corresponds to the photon sphere radius $r_m$ and the critical impact parameter $b_c$, given by:
\begin{align}
\label{geo_bc}
\frac{1}{b_c^2} &= \frac{\mathrm{A}_{m}(r,\Theta,\lambda)}{\mathrm{D}_{m}(r,\Theta,\lambda)} \notag \\
&= \left[1 - \frac{\Theta^2 M (11M - 4r_m)}{2(\lambda+1) r_m^4} - \frac{2M}{r_m}\right] \left[r_m^2 - \frac{\Theta^2 \left(-4(\lambda+1) M r_m + (\lambda+1) r_m^2 + 2M^2\right)}{4(\lambda+1) r_m (2M - r_m)}\right]^{-1},
\end{align}
where the subscript $m$ denotes evaluation at $r = r_m$. Furthermore, Eq. \eqref{geo1_re2} reduces at $r = r_m$ to:
\begin{align}
\label{geo1_m}
\frac{\mathrm{D}'_m(r,\Theta,\lambda)}{\mathrm{D}_{m}} - \frac{\mathrm{A}'_m(r,\Theta,\lambda)}{\mathrm{A}_{m}} = 0.
\end{align}
To properly analyze photon trajectories, we reformulate the equation of motion using Eqs. \eqref{dot_p} and \eqref{dot_r} as:
\begin{align}
\left(\frac{\mathrm{d}r}{\mathrm{d}\varphi}\right)^2 + V(r) = 0,
\end{align}
where the effective potential for photons is:
\begin{align}
\label{geo_potential}
V(r) = \frac{\mathrm{D}(r,\Theta,\lambda)}{\mathrm{B}(r,\Theta,\lambda)} R(r),
\end{align}
with
\begin{align}
\label{geo_R}
R(r) \equiv \frac{\mathrm{D}(r,\Theta,\lambda)}{b^2 \mathrm{A}(r,\Theta,\lambda)} - 1.
\end{align}

The circular photon orbits at $r = r_m$ and their corresponding critical impact parameters $b_c$ are determined numerically by simultaneously solving $V(r) = 0$ and $V'(r) = 0$. In the strong deflection regime, the bending of light is described by the angle $\alpha(r_0)$. This angle characterizes the total deviation experienced by a photon that travels in from infinity, reaches its minimum radial distance $r = r_0$, and then returns outward to infinity. In other words, it is given by
\begin{align}
\label{geo_alpha}
\alpha(r_0) = I(r_0) - \pi,
\end{align}
with
\begin{align}
\label{geo_I}
I(r_0) \equiv 2 \int_{r_0}^{\infty} \left[\frac{\mathrm{D}(r,\Theta,\lambda)R(r)}{\mathrm{B}(r,\Theta,\lambda)}\right]^{-1/2}  \mathrm{d}r.
\end{align}
It is important to mention that the asymptotic area where the circumferential radius $R$ approaches infinity is shown here by the symbol $r_\infty$. Notably, $r_0$ has to meet the requirement that $V(r_0) = 0$, which suggests that the effective potential disappears at the turning point. The correlation between $r_0$ and $b$ is inherent in this condition. From now on, any values that have the subscript $0$ are evaluated on the equatorial plane at $r = r_0$.

We focus on the strong deflection limit (SDL), where the turning point $r_0$ approaches the photon sphere radius $r_m$. To isolate the features of the logarithmic divergence in this regime, we introduce a new radial coordinate:
\begin{align}
z \equiv 1 - \frac{r_0}{r},
\end{align}
so that $z = 0$ corresponds to the point of closest approach. This substitution yields a coordinate-independent formulation of the strong deflection limit. The integral $I(r_0)$ can then be expressed as:
\begin{align}
I(r_0) \equiv \int_0^1 f(z, r_0)  \mathrm{d}z, \quad \text{where} \quad f(z, r_0) \equiv \frac{2r_0}{\sqrt{H(z, r_0)}},
\end{align}
with
\begin{align}
H(z, r_0) \equiv \frac{\mathrm{D}(r,\Theta,\lambda) R(r)}{\mathrm{B}(r,\Theta,\lambda)}(1 - z)^4.
\end{align}
Expanding $H(z, r_0)$ in a power series in $z$:
\begin{align}
H(z, r_0) = \sum_{n=0}^\infty c_n(r_0) z^n,
\end{align}
we find the leading coefficients satisfy $c_0(r_0) = 0$ and, at the critical radius $r_0 = r_m$, also $c_1(r_m) = 0$. The coefficient $c_2(r_m)\neq0$ and takes the form:
\begin{align}
c_2(r_m) = \frac{r_m^2 \mathrm{D}_{m}(r,\Theta,\lambda)}{2 \mathrm{B}_{m}(r,\Theta,\lambda)} \left( \frac{\mathrm{D}_{m}''(r,\Theta,\lambda)}{\mathrm{D}_{m}(r,\Theta,\lambda)} - \frac{\mathrm{A}_{m}''(r,\Theta,\lambda)}{\mathrm{A}_{m}(r,\Theta,\lambda)} \right).
\end{align}
It guarantees that $H(z, r_m) \simeq c_2(r_m) z^2$ is close to the photon sphere. We divide $I(r_0)$ into a regular portion $I_R(r_0)$ and a divergent part $I_D(r_0)$ in order to extract the divergent behavior, where:
\begin{align}
I_D(r_0) \equiv \int_0^1 \frac{2r_0}{\sqrt{c_1(r_0) z + c_2(r_0) z^2}}  \mathrm{d}z.
\end{align}

Using the expansions:
\begin{align}
c_1(r_0) &\simeq \frac{r_m \mathrm{D}_{m}(r,\Theta,\lambda)}{\mathrm{B}_{m}(r,\Theta,\lambda)} \left( \frac{\mathrm{D}_{m}''(r,\Theta,\lambda)}{\mathrm{D}_{m}(r,\Theta,\lambda)} - \frac{\mathrm{A}_{m}''(r,\Theta,\lambda)}{\mathrm{A}_{m}(r,\Theta,\lambda)} \right)(r_0 - r_m), \\
b(r_0) &\simeq b_c + \frac{1}{4} \sqrt{\frac{\mathrm{D}_{m}(r,\Theta,\lambda)}{\mathrm{A}_{m}(r,\Theta,\lambda)}} \left( \frac{\mathrm{D}_{m}''(r,\Theta,\lambda)}{\mathrm{D}_{m}(r,\Theta,\lambda)} - \frac{\mathrm{A}_{m}''(r,\Theta,\lambda)}{\mathrm{A}_{m}(r,\Theta,\lambda)} \right)(r_0 - r_m)^2,
\end{align}
the divergent integral evaluates to:
\begin{align}
I_D(b) = -\frac{r_m}{\sqrt{c_2(r_m)}} \ln\left( \frac{b}{b_c} - 1 \right) + \frac{r_m}{\sqrt{c_2(r_m)}} \ln\left( r_m K_m \right),
\end{align}
where the finite equation $K_m \equiv \mathrm{D}_{m}''/\mathrm{D}_{m} - \mathrm{A}_{m}''/\mathrm{A}_{m}$ depends on $M$, $\Theta$, $\lambda$, and $r_m$. The standard portion is provided by:
\begin{align}
I_R(r_0) = \int_0^1 \left[ \frac{2r_m}{\sqrt{H(z, r_m)}} - \frac{2r_m}{\sqrt{c_1(r_m) z + c_2(r_m) z^2}} \right] \mathrm{d}z,
\end{align}
and remains finite as $r_0 \to r_m$. This integral must be evaluated numerically. In the strong deflection limit, the total deflection angle exhibits a logarithmic divergence of the form:
\begin{align}
\alpha(b) = -\bar{a} \ln\left( \frac{b}{b_c} - 1 \right) + \bar{b},
\end{align}
where the coefficients are:
\begin{align}
\bar{a} &= \sqrt{ \frac{2 \mathrm{A}_{m}(r,\Theta,\lambda) \mathrm{B}_{m}(r,\Theta,\lambda)}{\mathrm{A}_{m}(r,\Theta,\lambda) \mathrm{D}_{m}''(r,\Theta,\lambda) - \mathrm{D}_{m}(r,\Theta,\lambda) \mathrm{A}_{m}''(r,\Theta,\lambda)} } = \sqrt{ -\frac{2}{V_m''} }, \\
\bar{b} &= \bar{a} \ln\left( r_m^2 K_m \right) + I_R(r_m) - \pi.
\end{align}
In this case, $\bar{b}$ includes finite contributions from the regular part of the integral, whereas $\bar{b}$ measures the divergence strength close to critical orbits.

\begin{figure}
\centering
\includegraphics[scale=0.6]{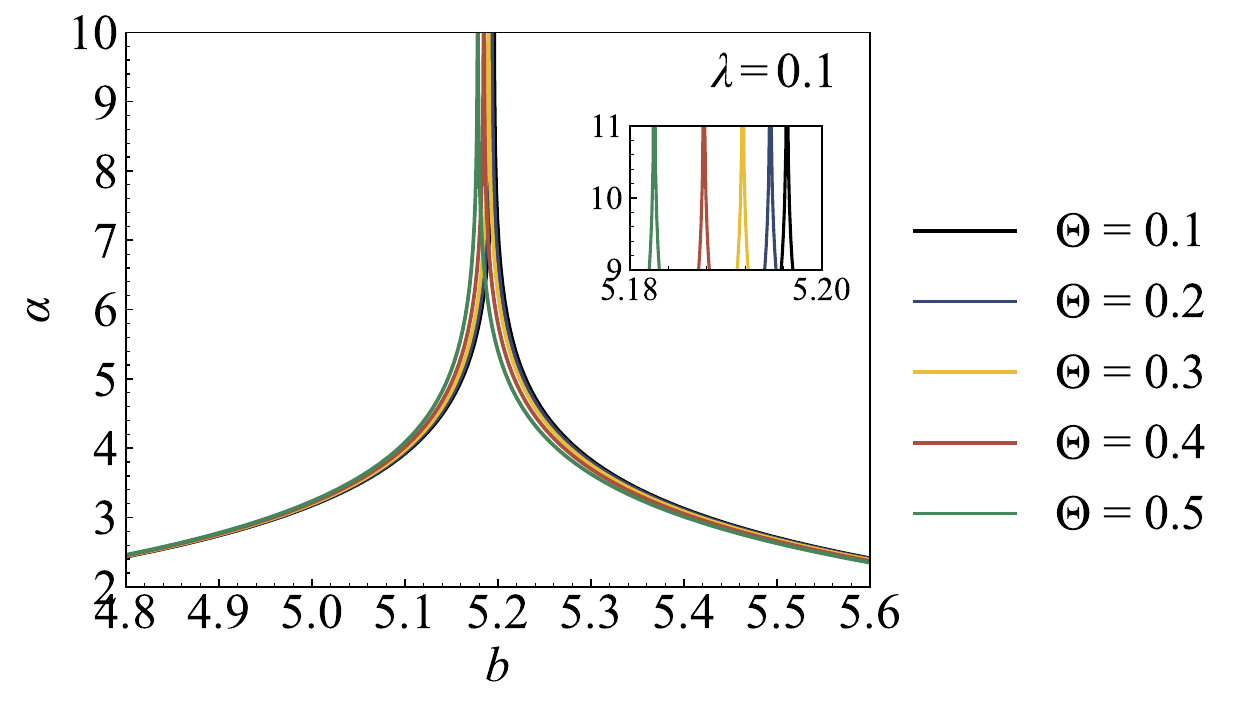}
\caption{The behavior of the deflection angle $\alpha$ in the strong--field regime is displayed for several choices of the parameter $\Theta$, keeping $\ell$ constant.}
\label{alpha_Theta}
\end{figure}

\begin{figure}
\centering
\includegraphics[scale=0.6]{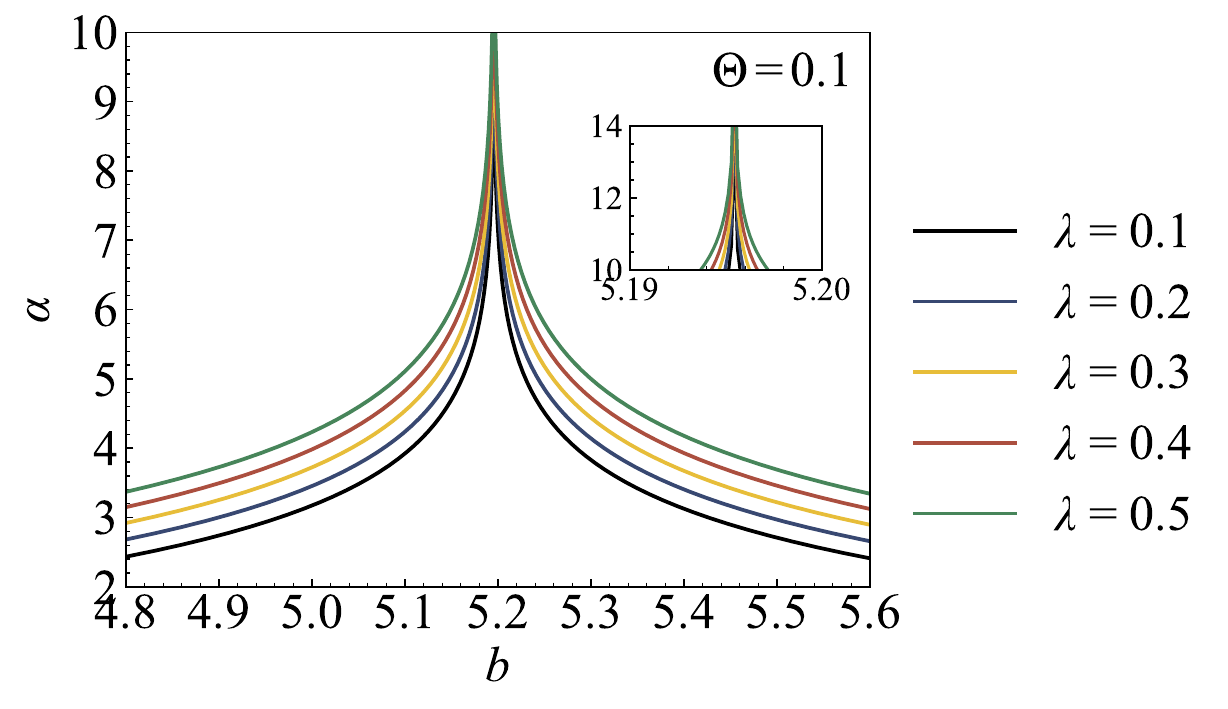}
\caption{In the strong--field limit, the deflection angle $\alpha$ is depicted as a function of the parameter $\lambda$, with $\Theta$ held fixed.}
\label{alpha_lambda}
\end{figure}

The strong gravitational deflection angle $\alpha$ is shown as a function of the impact parameter $b$ in Fig. \ref{alpha_Theta} for a range of non--commutative parameter $\Theta$ values, while holding $\lambda$ constant. The entire curve moves towards lower values of $b$ as $\Theta$ rises, indicating an improvement in the metric's non--commutative corrections. These adjustments decrease the effective gravitational focussing by introducing more departures from the Schwarzschild scenario. The steady peak location emphasises the photon sphere's stability when parameters are changed, and the curve's leftward shift shows how non--commutativity affects gauge singularity behaviour. The delicate structure of the deflection near the critical area is visible under a magnified view, indicating its acute and sensitive character, which is in line with the metric formalism's expectations.

With $\Theta$ set at $0.1$, Fig. \ref{alpha_lambda} shows how the deflection angle $\alpha$ depends on the impact parameter $b$ over a range of $\lambda$ values. The apex of the deflection angle is close to $b \approx 5.196$. The photon sphere in the conventional Schwarzschild metric is tightly aligned with a prominent peak at $b \approx 5.196$, which is present in all curves that correspond to $\lambda$ values between $0.1$ and $0.5$. This implies that the crucial impact parameter is maintained even when the non--commutative correction modifies the deflection's magnitude. Larger $\lambda$ values suppress the non--commutative corrections in the metric components, especially in $\mathrm{A}(r,\Theta,\lambda)$ and $\mathrm{B}(r,\Theta,\lambda)$, where $\lambda$ enters inversely, making the spacetime closer to Schwarzschild and intensifying the gravitational lensing. This is demonstrated by the deflection angle increasing significantly with $\lambda$. The curves in the inset show exponential convergence inside $b \in [5.19, 5.20]$, suggesting a universal scaling behaviour close to the crucial area.\\


\section{Bounds of Non-Commutativity via EHT measurements}
Data collected by the Event Horizon Telescope (EHT) on the black hole $Sgr A^*$ \cite{akiyama2022firstSgr,wielgus2022millimeter,akiyama2022firstSgrA} provide an important testbed for probing departures from general relativity \cite{araujo2025optical,araujo2025gravitationalfRT}. In this work, we employ the shadow data of $Sgr A^*$ to constrain the bumblebee parameter $\lambda$ in the presence of spacetime noncommutativity characterized by $\Theta$.

The key observable is the shadow's angular diameter, $\Omega_{\text{sh}}$. It is directly related to the impact parameter and the distance to the observer $D_O$ as follows \cite{afrin2023tests,kumar2020rotating}
\begin{equation}
\Omega_{\text{sh}} = \frac{2b_c}{D_{O}}.
\end{equation}
In observational units, this becomes \cite{xu2025optical,heydari2024effect}
\begin{equation}
\Omega_{\text{sh}} = \frac{6.191165 \times 10^{-8}\gamma}{\pi{D_{O}/{\text{Mpc}}}} \frac{b_c}{M}(\mu\text{as}),
\end{equation}
with $\gamma = M/M_\odot$.

For $Sgr A^*$, the EHT collaboration reports $M = 4 \times 10^6 M_\odot$, $D_{O} = 8.15$ kpc, and a shadow size of $\Omega_{\text{sh}} = 48.7 \pm 7 (\mu\text{as})$ \cite{akiyama2022firstSgr,wielgus2022millimeter,akiyama2022firstSgrA}. These values define the empirical window against which theoretical predictions are confronted.

\begin{figure}[ht]
\centering
\includegraphics[width=90mm]{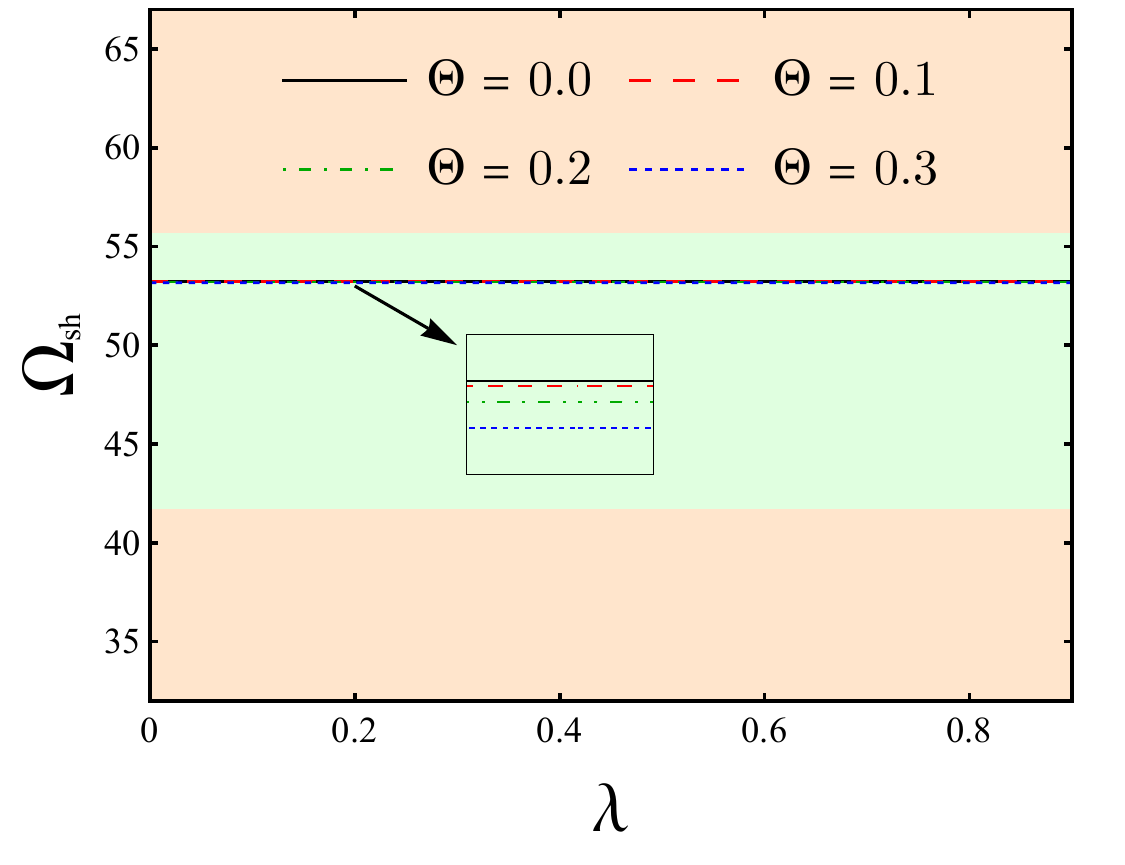}
\caption {The angular diameter of the black hole shadow ($\Omega_{\text{sh}}$) is plotted against the bumblebee gravity parameter ($\lambda$), for fixed $\Theta$. The green region indicates the EHT observational range.}
\label{fig:SgrALambda}
\end{figure}

\begin{figure}[ht]
\centering
\includegraphics[width=90mm]{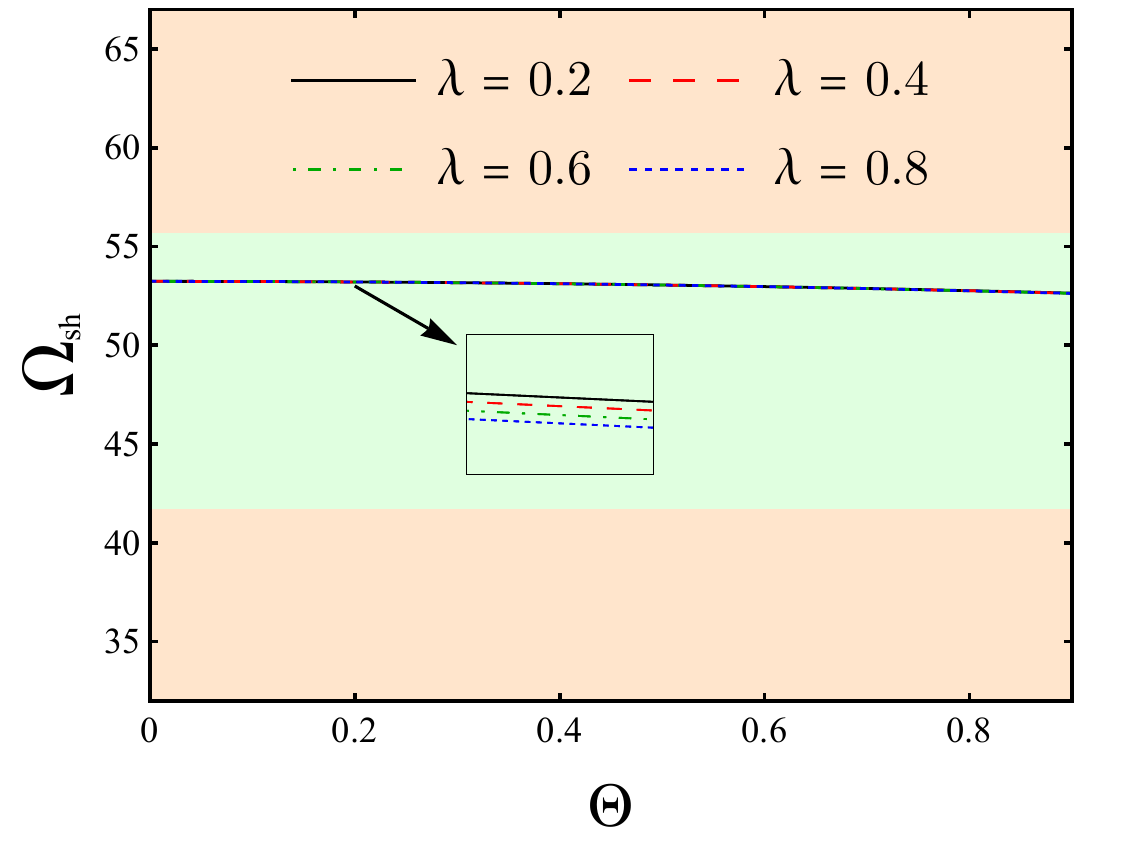}
\caption{Dependence of angular shadow diameter $\Omega_{\text{sh}}$ on the non--commutative parameter $\Theta$ for different $\lambda$. The EHT measurement of $Sgr A^*$'s angular shadow diameter is indicated by the green band.}
\label{fig:SgrATheta}
\end{figure}

Figure~\ref{fig:SgrALambda} shows the dependence of the shadow angular diameter $\Omega_{\text{sh}}$ on the bumblebee parameter $\lambda$ for different fixed values of the non--commutative parameter $\Theta$. The variation with respect to $\lambda$ is relatively small; however, the shadow angular diameter remains within the EHT observational range for $Sgr A^*$ across the considered values of $\Theta$. 
Figure~\ref{fig:SgrATheta} represents the effect of varying the non--commutative parameter $\Theta$ on the angular diameter of shadow for different values of the bumblebee parameter. Increasing $\Theta$ leads to a mild decrease in $\Omega_{\text{sh}}$ across all choices of $\lambda$. It is also evident that a wide range of $\Theta$ values produces shadow angular diameters that remain consistent with the EHT measurement of $48.7 \pm 7~\mu\text{as}$.




\section{\label{Sec12}Bounds Derived from Solar System Observations}

The classical tests from the solar system are considered the first verifications of the validity of General Relativity (GR) as a theory of gravity as geometrical manifestation of the spacetime curvature. Until today these observations are enhanced and serve as bases to test possible departures from GR.

In a straightforward way, we consider that the geometry of the solar system environment is given by the exterior manifold generated by a source mass with time and spherical symmetry, which is the gravitational field of the Sun, which for the GR case is the Schwarzschild solution, but in our case is our black hole metric. Massive test particles are planets like Mercury, which follow timelike geodesics and light-rays that reach the Earth are null geodesics.

To study such motion, we consider the simplest case where $\theta = \pi/2$ (confined in the equatorial plane) in our coordinates, and the Lagrangian that rules the motion is given by
\begin{equation}\label{eq:lagr1}
\mathrm{A}(r,\Theta,\lambda) \, {{\dot t}^2} - \mathrm{B}(r,\Theta,\lambda)\,{{\dot r}^2} - \mathrm{D}(r,\Theta,\lambda)\,{{\dot \varphi }^2} = \eta\, .
\end{equation}

The parametrization and nature (whether timelike or null) of these curves is determined by a parameter $\eta$ such that $L(x, \dot{x}) = -\eta/2$. Where $\eta=0$ describes massless particles, while, on the other hand, $\eta = 1$ describes to massive constituents. Due to the symmetries of this spacetime, we have two conserved charges, associated $E$ and $L$ of the particles, as commented in the previous sections, we have
\begin{equation}\label{constant2}
E = \mathrm{A}(r,\Theta,\lambda) \,\dot t \quad\mathrm{and}\quad L = \mathrm{D}(r,\Theta,\lambda)\,\dot \varphi.
\end{equation}

If we manipulate Eq.~\eqref{eq:lagr1} with the expressions \eqref{constant2}, we can derive the useful expression below
\begin{equation}\label{massive}
    \left[\frac{\mathrm{d}}{\mathrm{d}\varphi}\left(\frac{1}{r}\right)\right]^2=r^{-4} \,\mathrm{D}^{2}(r,\Theta,\lambda)\left[\frac{E^2}{\mathrm{A}(r,\Theta,\lambda)\mathrm{B}(r,\Theta,\lambda)L^2}-\frac{1}{\mathrm{B}(r,\Theta,\lambda)L^2}\left(\eta+\frac{L^2}{\mathrm{D}(r,\Theta,\lambda)}\right)\right]\, .
\end{equation}

Here, let us also differentiate Eq.~\eqref{massive} with respect to $\varphi$ to derive a second order differential equation. For our purposes, an interesting variable is $u = L^2/(M r)$ which allows us to rewrite our equation of motion \eqref{massive}. Using the metric functions \eqref{gtt}-\eqref{gphi}, our test particles obey the following geodesic equation up to first order in $\lambda$ and second order in $\Theta$:
\begin{align}\label{eq:u-massive}
&\frac{\mathrm{d}^2 u}{\mathrm{d}\varphi^2} = \left(1-\lambda\right)\left(\eta - u + \frac{3 M^2 u^2}{L^2}\right)\nonumber\\
&+\frac{\Theta^2}{2 L^8 (L^2 - 2 M^2 u)^2} \Bigg(L^{10} E^2 u-L^{10} \eta  u-10 L^8 E^2 M^2 u^2+7 L^8 \eta  M^2 u^2-L^8 M^2 u^3\nonumber\\
&+32 L^6 E^2 M^4 u^3-14 L^6 \eta  M^4 u^3+4 L^6 M^4 u^4-30 L^4 E^2 M^6 u^4+4 L^4 \eta  M^6 u^4+2 L^4 M^6 u^5\nonumber\\
&+8 L^2 \eta  M^8 u^5-24 L^2 M^8 u^6+24 M^{10} u^7 \Bigg).
\end{align}

Notice that the bumblebee parameter acts as a conformal factor to the relativistic term in the first line of \eqref{eq:u-massive}, while the non--commutative contribution couples with the energy of the particles in a way that its effect would be enhanced for very energetic test bodies. In fact, this is a property that is usually found in models of quantum gravity \cite{Amelino-Camelia:2008aez,Addazi:2021xuf,AlvesBatista:2023wqm}.

Also, it is important to highlight that the Newtonian contribution shows up in the first line. The GR correction involves the factor $M^2/L^2$, or recovering Newton's constant and the speed of light, $G^2M^2/(c^4L^2)$, which is a term that is much smaller than the Newtonian one, as it is suppressed by the speed of light. Further expanding this expression in terms of powers of the $M/L$ contribution leads to
\begin{align}\label{eq:u}
     u''(\varphi)= \, (1-\lambda)\eta -\frac{u \left(2 L^4 (1-\lambda)+L^2 \Theta ^2 \left(-E^2+\eta\right)\right)}{2 L^4}+\frac{3 M^2 u^2 \left(2(1-\lambda) L^2+\Theta ^2 \left(-2E^2+\eta \right)\right)}{2 L^4}\, .
\end{align}

We shall work with this perturbative scenario to analise the classical tests of the solar system and estimate the magnitude that the non--commutative and local Lorentz violating effects can have to not contradict the observations.


\subsection{Mercury’s Orbital Precession}

The correct prediction of Mercury's perihelion precession is one of the great successes of General Relativity, which is measured as angular shift in arcseconds per century. Mercury is described as massive test body (therefore $\eta=1$) of $E$ and $L$ in the gravitational field of the Sun of mass $M$. For convenience, we introduce a redefined mass $m=M/L$, from which the trajectory equation \eqref{eq:u} becomes 
\begin{equation}\label{eq:merc}
     u''(\varphi)= \, 1-\lambda -u+\alpha u+3 m^2 u^2\, .
\end{equation}
where the parameter $\alpha$ is defined as $\alpha = \lambda - \frac{\Theta^2}{2L^2}(1 - E^2)$. Since the bumblebee term only sums up to the equation, we can express the solution perturbatively in terms of $m$ and $\alpha$, adopting the expansion $u = u_0 + \alpha u_{\alpha}+m^2 u_m$. In this expression, the the Newtonian solution is given by $u_0$ plus a correction that depends on $\lambda$, explicitly written as:
\begin{equation}
    u_0=1-\lambda +e \cos(\varphi)\, .
\end{equation}

This means that, in an isolated way, the bumblebee term corrects the Newtonian contribution, independently from the GR correction.

Now, if we insert the approximate form of $u$ into Eq.~\eqref{eq:merc}, we derive the following:
\begin{equation}\label{eq:pert_hay_merc}
    m^2 \left(3 (e \cos (\varphi )+1)^2-u_m''-u_m\right)+\alpha \left(1-\lambda+e \cos (\varphi )-u_{\alpha}''-u_{\alpha}\right)=0\, .
\end{equation}

If we do not consider contributions of order $m^2\alpha$ and $m^2\lambda$, since these are small parameters, the term that remains involving $m^2$ is the one predicted by GR. In this manner, it reads:
\begin{equation}
    u_m=3m^2\left[\left(1+\frac{e^2}{2}\right)-\frac{e^2}{6}\cos\left(2\varphi\right)+e\varphi\sin\left(\varphi\right)\right]\, .
\end{equation}

Notice that terms that simply oscillate and those that are constant do not contribute to the shift in the perihelion, which means that in can exclude them in the final form of the solution that will be used to calculate the shift. On the other hand, the terms that are proportional to $\varphi$ indeed grow with each orbital cycle, which furnishes an amplification that accumulates and becomes observable over time.

The term proportional to $\alpha$ in Eq.~\eqref{eq:pert_hay_merc} must vanish, which can be solved to give the perturbed solution $u_{\alpha}(\varphi)$. This function has various oscillating terms that average around zero and do not contribute to the perihelion advance. The only term that indeed contributes to the effect is given by $-\frac{1}{2}e\varphi \sin(\varphi)$. Considering the usual and the new terms, the expression for $u(\varphi)$ that will be used to derive the advance is given by
\begin{equation}
    u(\varphi)=1-\lambda +e \cos (\varphi )+\frac{3M^2}{L^2}\left(1+\alpha\frac{L^2}{6M^2}\right)e\varphi\sin(\varphi)\, .
\end{equation}

In order to further simplify this expression, since the contribution from $\varphi\sin(\varphi)$ has no impact due to the smallness of the perihelion advance, in the final expression, the remaining two terms should be merged by taking into account some trigonometric identities of the cosine function, therefore, giving:
\begin{equation}
    u(\varphi)\approx 1-\lambda+e\cos\left[\left(1-\frac{3M^2}{L^2}\left(1+\alpha\frac{L^2}{6M^2}\right)\right)\varphi\right]\doteq 1-\lambda+e\cos\left[\left(1-\frac{3\widetilde{M}^2}{L^2}\right)\varphi\right]\, .
\end{equation}

Technically, this result can be interpreted as a redefinition of the mass parameter as $\widetilde{M}^2 = M^2 \left(1 +\frac{\alpha L^2}{6M^2} \right)$, which reflects in a modification of the GR description expressed by the mass $\widetilde{M}$ instead of $M$. From this, we verify that the extra addition to the precession of the perihelion is given by
\begin{equation}
    \Delta\Phi=6\pi\frac{\widetilde{M}^2}{L^2}=6\pi\frac{M^2}{L^2}\left(1+\alpha\frac{L^2}{6M^2}\right)\, .
\end{equation}

Therefore, the deviation from the General Relativity (GR) result in a dimensionless form is $\delta_{\text{Perih}} = \frac{\alpha L^2}{6M^2}$. Mercury's orbital period is nearly 88 days, leading to roughly 415 revolutions each century, calculated as $100 \times 365.25 / 88 \approx 415$. The cumulative precession over a century is found by multiplying this number of orbits by the angular shift per orbit. The GR prediction for this centennial precession is $\Delta \Phi_{\text{GR}} = 42.9814''$, a value that is consistent with the observational result $\Delta \Phi_{\text{Exp}} = (42.9794 \pm 0.0030)''$/century \cite{Casana:2017jkc,Yang:2023wtu}. This agreement supports the theory. The experimental uncertainty, however, defines a range within which potential contributions from alternative gravitational theories, parameterized by $\lambda$ and $\Theta$, can be investigated.

The quantity $L$ for a planetary orbit is related to its semi-major axis $a$ and eccentricity $e$ by $L^2 = M a (1 - e^2)$. Meanwhile, the orbital energy per unit mass is given by $E = -M/(2a)$ \cite{Goldstein:2002}. For Mercury's orbit around the Sun, the relevant parameters in natural units are: orbital eccentricity $e = 0.2056$, solar mass $M = M_{\odot} = 9.138 \times 10^{37}$, and orbital semi-major axis $a = 3.583 \times 10^{45}$. Inputting these values yields $L = 5.600 \times 10^{41}$. The smallness of the term $M^2/L^2$ justifies the use of a perturbative method. Furthermore, with a calculated squared energy of $E^2 = 1.627 \times 10^{-16}$, which is negligible, any energy-dependent corrections can be safely ignored.

Using the parameter $\alpha = \lambda - \frac{\Theta^2}{2L^2}(1 - E^2)$, we can derive constraints on $\lambda$ and $\Theta$, independently. If we set $\Theta = 0$, we isolate the Bumblebee contribution so that it leads to the following constraint $-1.817 \times 10^{-11} \leq \lambda \leq 3.634 \times 10^{-12}$. In an alternative way, if we take $\lambda = 0$, it allows to set bounds on the non--commutative parameter, resulting in {$\Theta^2\leq 1.14\times 10^{73}$ in natural units or $\Theta^2 \leq 2976.57\, \text{m}^2$ in S. I. . In a dimensionless way, we can compare this quantity with the solar length scale $M = GM_\odot/c^2 = 1.48$ km as $\Theta^2/M^2 \leq 1.36\times 10^{-3}$.}


\subsection{Light Bending}

Whenever a light ray passes just near the surface of a massive object, its path is bended due to the spacetime curvature, which modifies the apparent position of the source of the light ray from the observer's point of view. Such bending can be considered by analyzing null geodesics in a given spacetime, like the one that we are studying in this paper. Setting $\eta=0$ in Eq.~\eqref{eq:u}, and by redefining $u=1/r$, we arrive at the following differential equation:
\begin{align}\label{eq:light}
    u''(\varphi)=&\frac{-2L^2(1-\lambda)+E^2\Theta^2}{2L^2}u+\frac{3 M \left(-2E^2 \Theta ^2-2L^2\lambda+ 2L^2\right)}{2L^2}u^2 -\frac{\Theta^2}{2}u^3\, .
\end{align}

An interesting observation is that the quantity $L/E$ defines the impact parameter $b$. To study in an isolated way the effects of the non--commutativity parameter $\Theta$, we first set $\lambda = 0$. Curiously, we have corrections that depend just on the parameter $\Theta$, without a coupling with the mass $M$. The expression that we will use in the analysis is the following:
\begin{equation}\label{eq:redef_light_theta}
     u''(\varphi)+\left(1-\frac{\Theta^2}{2b^2}\right)u=3\widetilde{M}u^2-\frac{\Theta^2}{2}u^3\, ,
\end{equation}
where we defined an effective mass as $\widetilde{M} = M\left(1 - \frac{\Theta^2}{b^2}\right)$ that will simplify our calculation.  The solution should be given by a deformed Newtonian term \cite{Yang:2023wtu} given by 
\begin{equation}
    u_0=b^{-1}\sin\left(\left(1-\frac{\Theta^2}{4b^2}\right)\varphi \right)\, ,
\end{equation}
which is found by considering that the left--hand side of this expressoin as zero, plus corrections due to the mass $\widetilde{M}$ and $\Theta$. The corrected Newtonian term does not present a shifted trajectory, and the perihelion advance should come from the other terms.

Now, inserting this result into Eq.~\eqref{eq:redef_light_theta} and applying the small-angle approximation ($\varphi \ll 1$), the trajectory equation becomes
\begin{equation}
u(\varphi)=\frac{1}{b}\sin\left(\left(1-\frac{\Theta^2}{4b^2}\right)\varphi\right)+\frac{\widetilde{M}}{b^2(1-\Theta^2/(2b^2))}\left[1+\cos^2\left(\left(1-\frac{\Theta^2}{4b^2}\right)\varphi\right)\right], .
\end{equation}

The asymptotic behavior of the photon, where $u \to 0$ (i.e., $r \to \infty$), determines the scattering angles. Applying this boundary condition and performing a perturbative expansion yields the angles $\varphi_{\text{in}} = -2\bar{M}/b$ and $\varphi_{\text{ex}} = \pi + 2\bar{M}/b$, with the effective mass defined as $\bar{M} = M\left(1 - \frac{\Theta^2}{4b^2}\right)$. The total deflection is then $\delta_{\Theta} = -2\varphi_{u \to 0}$, leading to:
\begin{equation}
\delta_{\Theta}=\frac{4\bar{M}}{b}=4\frac{M}{b}\left(1-\frac{\Theta^2}{4b^2}\right), .
\end{equation}

For a light ray grazing the Sun, the impact parameter is $b \simeq R_{\odot} = 4.305 \times 10^{43}$, and the solar mass is $M = M_{\odot} = 9.138 \times 10^{37}$. In this scenario, the non--commutative correction to the deflection is encapsulated in the factor $1 - \Theta^2/(4b^2)$.

The General Relativity prediction is $\delta_{\text{GR}} = 4M/b = 1.7516687''$. The experimental measurement is expressed as $\delta_{\text{Exp}} = \frac{1}{2}(1 + \gamma) \times 1.7516687''$, where $\gamma = 0.99992 \pm 0.00012$ \cite{dsasdas}. To constrain $\Theta$, the theoretical factor $1 - \Theta^2/(4b^2)$ is equated to the experimental ratio $(1 + \gamma)/2$, resulting in the bounds: {$\Theta^2\leq 1.936\times 10^{14}\, \text{m}^2$ or $\Theta^2\leq 7.413\times 10^{83}$ in natural units, or comparing with the solar length scale $M = GM_\odot/c^2 = 1.48$ km as $\Theta^2/M^2=8.839\times 10^7$.}

To isolate the Bumblebee field's effect, we set $\Theta = 0$ in Eq.~\eqref{eq:light}, simplifying the trajectory equation to:
\begin{equation}\label{eq:light-hay}
u''(\varphi)+(1-\lambda) u=3(1-\lambda) M u^2 =3{\cal M} u^2, .
\end{equation}
Here, ${\cal M}=(1-\lambda) M$ acts as an effective mass. The solution to this equation is
\begin{equation}
u(\varphi)=\frac{1}{b}\sin\left(\left(1-\frac{\lambda}{2}\right)\varphi\right)+\frac{{\cal M}}{b^2(1-\lambda)}\left[1+\cos^2\left(\left(1-\frac{\lambda}{2}\right)\varphi\right)\right], .
\end{equation}

Employing the small-angle approximation and the same asymptotic analysis, the deflection angle attributed to the Bumblebee field is found to be $\delta_{\lambda} = \frac{4{\cal M}}{b}(1 + \frac{3}{2}\lambda)\approx \frac{4M}{b}(1 + \frac{1}{2}\lambda)$. This result leads to the constraint on the parameter $\lambda$: $-2 \times 10^{-4} \leq \lambda \leq 4 \times 10^{-5}$.


\subsection{Time Delay (Shapiro Effect)}

The Shapiro time delay~\cite{Shapiro:1964uw} is the observable increase in the round-trip travel time of radar signals propagating to a planet and back to Earth, resulting from the curvature of spacetime induced by the Sun's gravitational field.

To compute the Shapiro delay, the propagation of light along null geodesics is examined, as described by Eq.~\eqref{massive}. By imposing the null condition $\mathrm{d}s^2 = 0$ and utilizing Eq.~\eqref{constant2}, we derive:
\begin{equation}
  \left(  \frac{\mathrm{d}r}{\mathrm{d}t}\right)^2=\frac{\mathrm{A}(r,\Theta,\lambda)\,\mathrm{D}(r,\Theta,\lambda)-\frac{L^2}{E^2}\mathrm{A}(r,\Theta,\lambda)}{\mathrm{B}(r,\Theta,\lambda)\,\mathrm{D}(r,\Theta,\lambda)}\, .
\end{equation}

As shown in Ref.~\cite{Wang:2024fiz}, the constants $E$ and $L$ can be related to the impact parameter $b$ of the light ray. The point of closest approach, $r_{\text{min}}$, is found by setting $\dot{r} = 0$, which implies $L^2/E^2 = \mathrm{D}(r_{\text{min}},\Theta,\lambda)/\mathrm{A}(r_{\text{min}},\Theta,\lambda)$. Consequently, the differential travel time is expressed as:
\begin{equation}\label{eq:shapiro_main}
    \mathrm{d} t=\pm \frac{1}{\mathrm{A}(r,\Theta,\lambda)}\frac{1}{\sqrt{\frac{1}{\mathrm{A}(r,\Theta,\lambda)\,\mathrm{B}(r,\Theta,\lambda)}-\frac{\mathrm{D}(r_{\text{min}},\Theta,\lambda)/\mathrm{A}(r_{\text{min}},\Theta,\lambda)}{\mathrm{B}(r,\Theta,\lambda)\,\mathrm{D}(r,\Theta,\lambda)}}}\, .
\end{equation}

We first consider the non--commutative geometry scenario by taking $\lambda = 0$. Retaining only the leading-order corrections in $M$ and $\Theta^2$, and integrating Eq.~\eqref{eq:shapiro_main}, we obtain:
\begin{align}
    t=\sqrt{r^2-r_{\text{min}}^2}+M\left(\sqrt{\frac{r-r_{\text{min}}}{r+r_{\text{min}}}}+2\ln\left(\frac{r+\sqrt{r^2-r_{\text{min}}^2}}{r_{\text{min}}}\right)\right)\\
    +\frac{\Theta ^2 \left(2 (r_{\text{min}}-M) \arctan\left(\frac{\sqrt{r^2-r_{\text{min}}^2}-r}{r_{\text{min}}}\right)-\frac{M (19 r_{\text{min}}+26 r) \sqrt{r^2-r_{\text{min}}^2}}{r (r_{\text{min}}+r)}\right)}{8 r_{\text{min}}^2}\, .\nonumber
\end{align}

In the limit $r \gg r_{\text{min}}$, the dominant terms from General Relativity and the leading non--commutative correction are:
\begin{equation}\label{eq:sh_t_nc}
    t(r)=r+M+2M\ln\left(\frac{2r}{r_{\text{min}}}\right)-\frac{13M}{4r_{\text{min}}^2}\Theta^2\, .
\end{equation}

Let $t(r_E)$ and $t(r_R)$ denote the one--way travel times from the emitter and receiver to the Sun, respectively, computed via Eq.~\eqref{eq:sh_t_nc}. The total round--trip time is then $T = 2t(r_E) + 2t(r_R)$, which evaluates to:
\begin{equation}
T_{\Theta}= 2(r_E+r_R)+4M\left[1+\ln\left(\frac{4r_Rr_E}{r_{\text{min}}^2}\right)-\frac{13\Theta^2}{4r_{\text{min}}^2}\right]=T_{\text{flat}}+\delta T\, .
\end{equation}

The Shapiro effect quantifies the additional travel time experienced by a signal as it moves through curved spacetime, relative to the propagation time in flat space, given by $T_{\text{flat}} = 2(r_E + r_R)$. Within the parametrized post--Newtonian (PPN) formalism, the corresponding relativistic contribution takes the form:
\begin{equation}
    \delta T = 4M\left(1+\frac{1+\gamma}{2}\ln \left(\frac{4r_Rr_E}{r_{\text{min}}^2}\right)\right)\, .
\end{equation}

High--precision measurements from the Cassini mission \cite{Bertotti:2003rm,Will:2014kxa} have established the tightest constraint on the PPN parameter $\gamma$, yielding $|\gamma - 1| < 2.3 \times 10^{-5}$. Using astronomical units, with $r_E = 1, \text{AU} = 2.457 \times 10^{45}$, $r_R = 8.46, \text{AU}$, and $r_{\text{min}} = 1.6 R_{\odot}$ where $R_{\odot} = 4.305 \times 10^{43}$, we derive an upper limit on the non--commutative parameter: {$\Theta^2 \leq 5.001 \times 10^{14}\, \text{m}^2$ or in natural units $\Theta^2\leq 1.915\times 10^{84}$. In a dimensionless way, we can express this result in units of the solar length scale $M = GM_\odot/c^2 = 1.48$ km, as $\Theta^2/M^2\leq 2.283\times 10^{8}$.}

For the Bumblebee model, we set $\Theta = 0$ and integrate Eq.~\eqref{eq:shapiro_main} to find the leading contributions:
\begin{align}
   t=\left(1+\frac{\lambda}{2}\right)\left[\sqrt{r^2-r_{\text{min}}^2}+M\left(\sqrt{\frac{r-r_{\text{min}}}{r+r_{\text{min}}}}+2\ln\left(\frac{r+\sqrt{r^2-r_{\text{min}}^2}}{r_{\text{min}}}\right)\right)\right]\, .
\end{align}

In the limit $b \ll r$, the asymptotic form becomes:
\begin{equation}
     t=\left(1+\frac{\lambda}{2}\right)\left[r+M+2M\ln\left(\frac{2r}{r_{\text{min}}}\right)\right]\, .
\end{equation}

The resulting Shapiro delay in the Bumblebee model is therefore:\begin{equation} T_{\lambda}=\left(1+\frac{\lambda}{2}\right)\left\{2(r_E+r_R)+4M\left[1+\ln\left(\frac{4r_Rr_E}{r_{\text{min}}^2}\right)\right]\right\}\, .
\end{equation}

Analysis of Cassini data constrains the Bumblebee parameter to $|\lambda| \leq 2.280 \times 10^{-5}$, using the solar mass $M$. 

A summary of the resulting constraints is presented in Table~\ref{tab:constr}.  {Non-commutative corrections involving massive particles are governed by corrections of the kind $\Theta^2/M^2$, while those involving massless particles are given by $\Theta^2/b^2$ or $\Theta^2/r_{\text{min}}^2$. Considering the solar system parameters, the massive correction is some orders of magnitude larger than the massless ones. Also, the dimensionless Bumblebee parameter $\lambda$ do not present any amplifier in the massless case, whereas in the massive one, it couples with the factor $L^2/M^2$ which amplifies the effect. These results are responsible for the derivation of better constraints from Mercury precession.}

\begin{table}[h!]
\centering
\caption{Bounds for $\Theta^2$ and $\lambda$  derived from Solar System tests.}
\label{tab:constr}
\begin{tabular}{lc}
\hline\hline
\textbf{Solar System Test} & Constraints ($M = GM_\odot/c^2 = 1.48$ km) \\
\hline
{\bf{Mercury precession}}   & \makecell{$\Theta^2/M^2 \leq 1.36\times 10^{-3}$ \\ $-1.817 \times 10^{-11} \leq \lambda \leq 3.634 \times 10^{-12} $} \\
{\bf{Light deflection}}     & \makecell{$\Theta^2/M^2\leq 8.839\times 10^7$ \\ $-2 \times 10^{-4} \leq \lambda \leq 4 \times 10^{-5}$}  \\
{\bf{Shapiro time delay}}   & \makecell{$\Theta^2/M^2\leq 2.283\times 10^{8}$ \\ $-2.280 \times 10^{-5}\leq \lambda \leq 2.280 \times 10^{-5}$}  \\
\hline\hline
\end{tabular}
\end{table}


\section{Conclusion}\label{Sec13}

In this work, we introduced a new black hole solution within the framework of bumblebee gravity, where spontaneous Lorentz symmetry breaking was driven by a vector field acquiring a non--vanishing vacuum expectation value. Non--commutative corrections were properly incorporated through the Moyal twist $\partial_{r} \wedge \partial_{\theta}$. Remarkably, the location of the event horizon $r_{h}$ remained unaffected by the presence of either the Lorentz--violating parameter $\lambda$ or the non--commutative parameter $\Theta$. Moreover, without expanding the geometric quantities---such as the Christoffel symbols and curvature tensors---the resulting solution proved to be regular. In particular, the Kretschmann scalar $K = R_{\mu\nu\rho\sigma}R^{\mu\nu\rho\sigma}$ was shown to approach a finite value in the limit $r \to 0$. In other words, it led to the absence of curvature singularities at the origin.

We then proceeded to investigate the propagation of light in this background by numerically solving a system of four coupled differential equations, which allowed us to trace in detail the behavior of null geodesics. In general lines, it was observed that increasing the Lorentz--violating parameter $\lambda$ caused the light trajectories to spread further apart, thereby modifying the deflection pattern of photons. The critical photon orbits $r_{c}$, obtained analytically as $r_{c} = 3M - \tfrac{\lambda \Theta^{2}}{9M}$, were subsequently confirmed to be unstable through an analysis of the associated Gaussian curvature of the optical metric and via topological method. Both parameters, $\lambda$ and $\Theta$, contributed to a reduction in the photon sphere radius $r_{c}$. To complement this analysis, we determined the shadow radius $\mathcal{R}$, which up to second order exhibited no dependence on $\lambda$, namely $\mathcal{R} \approx 3\sqrt{3}M - \tfrac{\Theta^{2}}{8\sqrt{3}M}$. As in the case of the photon sphere, larger values of the non--commutative parameter $\Theta$ led to smaller shadow radii.

Gravitational lensing was analyzed in both the weak-- and strong--deflection regimes, allowing us to probe the influence of the Lorentz--violating and non--commutative parameters on observable features. In the weak field limit, we computed the Gaussian curvature of the optical metric and extracted the deflection angle $\tilde{\alpha}(b,\Theta,\lambda)$. It was observed that increasing the non--commutative parameter $\Theta$ enhanced the deflection angle, whereas larger values of the Lorentz--violating parameter $\lambda$ suppressed it, in agreement with the qualitative behavior obtained from the analysis of null geodesics. For the strong deflection regime, we employed Igata’s method to highlighted the logarithmic divergence of the deflection angle near the photon sphere. Beyond this divergence point, the effect of the parameters was reversed: larger $\Theta$ values decreased the deflection angle, while increasing $\lambda$ amplified it. Moreover, the lensing observables were evaluated by confronting the theoretical predictions with Event Horizon Telescope (EHT) data for $Sgr A^{*}$ and $M87^{*}$.

Finally, we placed observational constraints on the parameters $\lambda$ and $\Theta$ by confronting the predictions of our model with high--precision Solar System experiments. From Mercury’s perihelion precession, we obtained the bounds $-595.315\,\text{m}^{2} \leq \Theta^{2} \leq 2976.57\,\text{m}^{2}$ and $-1.817 \times 10^{-11} \leq \lambda \leq 3.634 \times 10^{-12}$. Independent limits were derived from light--deflection measurements, leading to $-3.872 \times 10^{13}\,\text{m}^{2} \leq \Theta^{2} \leq 1.936 \times 10^{14}\,\text{m}^{2}$ and $-2 \times 10^{-4} \leq \lambda \leq 4 \times 10^{-5}$. In addition, Shapiro time--delay observations provided the broader interval $-5.001 \times 10^{14}\,\text{m}^{2} \leq \Theta^{2} \leq 5.001 \times 10^{14}\,\text{m}^{2}$ together with $-2.280 \times 10^{-5} \leq \lambda \leq 2.280 \times 10^{-5}$.

As a prospective direction, it would be of significant interest to derive a black hole solution directly from the action by incorporating non--commutative corrections together with the bumblebee field, in close analogy to the construction recently achieved in Ref.~\cite{Pozar:2025yoj}. Such an approach would allow for a more fundamental comparison with the bumblebee black hole solution presented in the present manuscript. In addition, within the formalism adopted here, the construction of a new black hole solution in bumblebee gravity under the metric--affine approach appears both feasible and particularly well--suited for further exploration. These possibilities of investigation, together with closely related developments, are currently being pursued by the same group of authors and will be reported in future work.


\acknowledgments{
A.A.A.F. is supported by Conselho Nacional de Desenvolvimento Cient\'{\i}fico e Tecnol\'{o}gico (CNPq) and Fundação de Apoio à Pesquisa do Estado da Paraíba (FAPESQ), project numbers 150223/2025-0 and 1951/2025. N. H. is supported by the Conselho Nacional de Desenvolvimento Científico e Tecnológico (CNPq), grant No. 152891/2025-0. I. P. L. was partially supported by the National Council for Scientific and Technological Development - CNPq grant 312547/2023-4.  I. P .L. would like to acknowledge networking support by the COST Action BridgeQG (CA23130) and the COST Action RQI (CA23115), supported by COST (European Cooperation in Science and Technology). N. H. would like to acknowledge networking support of the COST Action CA 22113 - Fundamental challenges in theoretical physics (Theory and Challenges), CA 21106 - COSMIC WISPers in the Dark Universe: Theory, astrophysics and experiments (CosmicWISPers), CA 21136 - Addressing observational tensions in cosmology with systematics and fundamental physics (CosmoVerse), and CA 23130 - Bridging high and low energies in search of quantum gravity (BridgeQG). Moreover, the authors would like to thank Filip Požar for the useful correspondence exchanged. FSNL acknowledges support from the Funda\c{c}\~{a}o para a Ci\^{e}ncia e a Tecnologia (FCT) Scientific Employment Stimulus contract with reference CEECINST/00032/2018, and funding through the research grants UIDB/04434/2020, UIDP/04434/2020 and PTDC/FIS-AST/0054/2021.}


	\bibliography{main}
	\bibliographystyle{ieeetr}
	
\end{document}